\documentstyle{article}

\def\hybrid{\topmargin 0pt      \oddsidemargin 0pt
        \headheight 0pt \headsep 0pt
        \textwidth 6.25in       % A4 paper
        \textheight 9.5in       % A4 paper
        \marginparwidth 0.0in
        \parskip 5pt plus 1pt   \jot = 1.5ex}
\catcode`\@=11
\def\marginnote#1{}
\newcount\hour
\newcount\minute
\newtoks\amorpm
\hour=\time\divide\hour by60
\minute=\time{\multiply\hour by60 \global\advance\minute by-\hour}
\edef\standardtime{{\ifnum\hour<12 \global\amorpm={am}%
        \else\global\amorpm={pm}\advance\hour by-12 \fi
        \ifnum\hour=0 \hour=12 \fi
        \number\hour:\ifnum\minute<10 0\fi\number\minute\the\amorpm}}
\edef\militarytime{\number\hour:\ifnum\minute<10 0\fi\number\minute}

\def\draftlabel#1{{\@bsphack\if@filesw {\let\thepage\relax
   \xdef\@gtempa{\write\@auxout{\string
      \newlabel{#1}{{\@currentlabel}{\thepage}}}}}\@gtempa
   \if@nobreak \ifvmode\nobreak\fi\fi\fi\@esphack}
        \gdef\@eqnlabel{#1}}
\def\@eqnlabel{}
\def\@vacuum{}
\def\draftmarginnote#1{\marginpar{\raggedright\scriptsize\tt#1}}

\def\draft{\oddsidemargin -.5truein
        \def\@oddfoot{\sl preliminary draft \hfil
        \rm\thepage\hfil\sl\today\quad\militarytime}
        \let\@evenfoot\@oddfoot \overfullrule 3pt
        \let\label=\draftlabel
        \let\marginnote=\draftmarginnote
   \def\@eqnnum{(\theequation)\rlap{\kern\marginparsep\tt\@eqnlabel}%
\global\let\@eqnlabel\@vacuum}  }

\def\numberbysection{\@addtoreset{equation}{section}
        \def\theequation{\thesection.\arabic{equation}}}

\def\underline#1{\relax\ifmmode\@@underline#1\else
        $\@@underline{\hbox{#1}}$\relax\fi}

\def\titlepage{\@restonecolfalse\if@twocolumn\@restonecoltrue\onecolumn
     \else \newpage \fi \thispagestyle{empty}\c@page\z@
        \def\thefootnote{\fnsymbol{footnote}} }

\def\endtitlepage{\if@restonecol\twocolumn \else  \fi
        \def\thefootnote{\arabic{footnote}}
        \setcounter{footnote}{0}}  %\c@footnote\z@ }
\relax

\numberbysection
\hybrid

\def\beq{\begin{equation}}
\def\eeq{\end{equation}}
\def\p{\partial}
\def\G{\Gamma}

\newtheorem{th}{Theorem}[section]
\newtheorem{prop}{Proposition}[section]
\newtheorem{cor}{Corollary}[section]

\def\square{\hfill
{\vrule height6pt width6pt depth1pt} \break \vspace{.01cm}}

\begin{document}

\begin{titlepage}

\title{Spin generalization of the Calogero-Moser system and the Matrix KP
equation}

\author{I. Krichever \thanks{ Landau Institute for Theoretical Physics
 Kosygina str. 2, 117940 Moscow, Russia}
\and O. Babelon $^{\dag} $ \and E. Billey $^{\dag} $  \and M. Talon
\thanks{L.P.T.H.E. Universit\'e Paris VI (CNRS UA 280), Box 126, Tour 16,
$1^{{\rm er}}$ \'etage, 4 place Jussieu, 75252 Paris Cedex 05, France}
}

\date{September 1994}

\bigskip

\maketitle

\begin{abstract}
The complete solutions of the spin generalization of the elliptic
Calogero Moser systems
are constructed. They are expressed in terms of Riemann theta-functions.
The analoguous constructions for the trigonometric and rational cases
are also presented.
\end{abstract}

\vfill
Preprint LPTHE 94/42 \hfill\break

\end{titlepage}
\newpage

\section{Introduction}

The elliptic Calogero-Moser system \cite{c},\cite{m} is a system of $N$
identical particles on a line interacting with each other via the
potential $V(x)=\wp(x)$, where $\wp(x)=\wp(x|\omega,\omega')$ is the
Weierstrass elliptic function with periods $2\omega,\ 2\omega'$.
This system (and its quantum version, as well) is a completely integrable
system \cite{op}. The complete solution of the elliptic Calogero--Moser
model was constructed by algebro--geometrical methods in \cite{kr2}.
The degenerate cases where $V(x) = 1/\sinh^2 x$ or $V(x)=1/x^2$ are also of
interest, and admit nice interpretations as reductions of geodesic motions
on symmetric spaces~\cite{op,AvBaTa93}. The analogous interpretation for the
elliptic case was recently given in~\cite{GorNe}.

In this work, we consider the spin generalization of the Calogero--Moser
model, which was defined in \cite{w}. Again this model exists
 in the elliptic, trigonometric and rational versions, each one
 being of its own interest. In particular the hidden symmetry of the model
 changes from a current algebra type in the rational case, to a yangian type
 in the trigonometric case~\cite{w,bghp,AvBaBi}.
Our main goal is to construct the action-angle type
variables for
these spin generalizations of the Calogero-Moser system, and to solve
the equations of motion in terms of Riemann theta-functions.
The algebro--geometric constructions
of the solutions  substantially differ in the three cases
and we shall present them in parallel.

Let us consider the classical hamiltonian system of $N$
particles on a line, with coordinates $x_i$ and momenta $p_i$, and internal
degrees of freedom
described for each particle by a  $l$-dimensional vector
$a_i=(a_{i,\alpha})$
and a $l$-dimensional co-vector $b_i^+=(b_i^{\alpha})$ , $\alpha=1,\ldots,l$.
The Hamiltonian has the form
\beq
H={1\over 2}\sum_{i=1}^N p_i^2 +{1\over 2} \sum_{i\neq
j}(b_i^+a_j)(b_j^+a_i)V(x_i-x_j).
\label{1}
\eeq
where  $(b_i^+a_j)$ stands for the corresponding scalar product
\beq
(b_i^+a_j)=(b_i^{\alpha}a_{i,\alpha}).\label{3}
\eeq
and the potential $V(x)$ is one of the functions $\wp(x), 1/\sinh^2 x$,
or $1/x^2$.
The non trivial Poisson brackets between the dynamical variables
$x_i,p_i,b_i^{\alpha},a_{i,\alpha}$ are
\beq
\{p_i,x_j\}=\delta_{ij},\ \
\{b_i^{\beta},a_{j,\alpha}\}=-\delta_{i,j}\delta_{\alpha}^{\beta}.\label{2}
\eeq
The equations of motion have the form
\beq
\ddot{x}_i=\sum_{j\neq i} (b_i^+a_j)(b_j^+a_i)V'(x_i-x_j), \ \
V'(x)={dV(x)\over dx}, \label{4}
\eeq
\beq
\dot{a}_i= -\sum_{j\neq i}a_j(b_j^+a_i)V(x_i-x_j),\label{5}
\eeq
\beq
\dot{b}_i^+=\sum_{j\neq i}b_j^+(b_i^+a_j)V(x_i-x_j)\label{6}
\eeq
 From (\ref{5},\ref{6}) it follows that $(b_i^+a_i)$ are integrals of
motions.
We restrict the system on the invariant submanifold
\beq
(b_i^+a_i)=c=2\label{7}
\eeq
{\bf Remark}. The reduction of the system (\ref{1}) onto the invariant
submanifold defined by the constraint (\ref{7}) is a
completely integrable hamiltonian system for any value of the constant
$c$. Changing the value of $c$ amounts to a rescaling of the time variable.
In the following we shall assume $c=2$ for definiteness.

Let us introduce the quantities
\beq
f_{ij}=(b_i^+a_j).\label{8}
\eeq
The Poisson brackets (\ref{2}) imply
\beq
\{f_{ij},f_{kl}\}=\delta_{jk}f_{il}-\delta_{il}f_{kj}. \label{9}
\eeq
The Hamiltonian (\ref{1}) in terms of these new variables has the form:
\beq
H={1\over 2}\sum_{i=1}^N p_i^2 + {1\over 2}
 \sum_{i\neq j}f_{ij}f_{ji}V(x_i-x_j)
\label{10}
\eeq
The system (\ref{10}) with $f_{ij}$ as dynamical variables satisfying the
relations (\ref{9}) was introduced in \cite{w} and was called
Euler--Calogero--Moser system. When $l<N$ the relations (\ref{8})
give a parametrisation of special symplectic leaves of the system
(\ref{9},\ref{10}).

Let us count the number of non--trivial
degrees of freedom. We start with
$2N+2Nl$ dynamical variables corresponding to the $x_i$, $p_i$,
$a_{i,\alpha}$,
$b_i^\alpha$. The Hamiltonian~(\ref{10}) has a symmetry under
rescaling:
\begin{equation}
a_i\to \lambda_i a_i, \quad b_i\to {1\over\lambda_i}b_i\label{local}
\end{equation}
(notice that $f_{ij}$ is non--invariant but $f_{ij}f_{ji}$ is
invariant, and the Poisson brackets are also invariant). The
corresponding moment is given by the collection of $b_i^+a_i$ and we
fix it to the values $b_i^+a_i=2$, which makes $N$ conditions.
The stabilizer of this moment consists in the whole group
so that the reduced system is defined by $N$ more constraints, e.g.
$\sum_\alpha b_i^\alpha =1$, leaving us with a
phase space of dimension $2Nl$. Moreover the Hamiltonian and the
symplectic structure are invariant under a further symmetry:
\begin{equation}
a_i\to W^{-1} a_i,\quad b_i^+ \to b_i^+ W \label{global}
\end{equation}
where $W$ is any matrix in $GL(r,{\bf R})$ independent of the label
$i$, preserving the above condition on the $b_i$'s. This means that
$W$ must leave the vector  $v=(1,\cdots,1)$ invariant. Hence this
group is of
dimension $l^2-l$. Fixing the momentum, $\cal P$, gives $l^2-l$
conditions. The
stabilizer of a generic momentum is trivial: this is because such a
generic element can be diagonalized as ${\cal P} = m^{-1}\Lambda m$. Its
stabilizer under the adjoint action consists of the matrices of the
form $g=m^{-1}Dm$ with $D$ diagonal. The condition $gv=v$ translates
into $D mv= mv$ wich implies generically $D=1$, i.e., $g=1$. We
have proved the

\begin{prop}
The dimension of the reduced phase space ${\cal M}$ is
\begin{eqnarray}
{\rm dim~} {\cal M} = 2\left[ Nl -{l(l-1) \over 2} \right]
\label{dim}
\end{eqnarray}
\end{prop}

Our method for the construction of the solutions of system (\ref{1})
is a generalization of the approach that was used for the classical
Calogero-Moser
system. In \cite{a} a remarkable connection between the Calogero-Moser system
and the motion of poles of the rational and the elliptic solutions of KdV
equation was found. It turned out that the corresponding relation becomes
an isomorphism in the case of the rational or the elliptic solutions of the
Kadomtsev-Petviashvili equation. In \cite{kr1} and \cite{chud}
this isomorphism in the rational case was used in
opposite directions.

In \cite{chud} using the known solutions of the rational
Calogero-Moser system, the rational solutions of KP equation were
constructed.

In \cite{kr1} the construction of rational solutions for various
partial differential equations admitting a zero-curvature representation
was proposed. Applying this result to the KP equation yielded an alternative
way to solve the Calogero-Moser system.
This approach was generalized in \cite{kr2} where
the action-angle variables for the elliptic Calogero-Moser system were
constructed
and the exact formula for elliptic solutions of KP equation was obtained.
(Further developments in the theory of so-called elliptic solitons are
presented in the special issue of Acta Applicandae Mathematicae
{\bf 35} (1994) dedicated to the memory of J.L. Verdier).

\section{Relation to the matrix KP equation.}

The zero-curvature representation for the KP equation
\beq
{3\over 4}u_{yy}=(u_t-{3\over 2}uu_x+{1\over 4} u_{xxx})_x\label{11}
\eeq
has the form (\cite{Z},\cite{dr})
\beq
[\p_y-L,\p_t-M]=0, \label{1.1}
\eeq
where
\beq
L=\p_x^2-u(x,y,t),\ \ \  M=\p_x^3-{3\over 2}u\p_x+w(x,y,t). \label{1.2}
\eeq

In this scalar case ($l=1$), assuming that $u$ is an elliptic function of
the variable $x$, the comparison of singular terms in the expansion of the
right and left hand sides of (\ref{11}) near the poles of $u$ gives directly
that:

1.) Any elliptic (in the variable $x$) solution of
KP equation
has the form
\beq
u(x,y,t)=2\sum_{i=1}^N \wp(x-x_i(y,t)) +const, \label{12}
\eeq

2.) The dependence of the poles $x_i(y,t)$ with respect to the variable
$y$ coincide with the elliptic Calogero-Moser system and their dependence
with respect to the variable $t$ is described by the ``third integral'' of this
system.

Let us consider the same equations (\ref{1.1}) in the case when operators
(\ref{1.2}) have matrix $(l\times l)$ coefficients. They are equivalent to
the system
\beq
w_x={3\over 4}u_y,\ \ w_y=u_t-{3\over 4}(uu_x+u_xu)+{1\over 4} u_{xxx}-[u,w],
\label{11a}
\eeq
that we call the matrix KP equation.

In the matrix case we don't know the complete classification of all elliptic
solutions of (\ref{11a}). It turns out that the system (\ref{1}) is isomorphic
to the special elliptic solutions of matrix KP equation having the form
\beq
u(x,y,t)=\sum_{i=1}^N \rho_i(y,t)\wp(x-x_i(y,t)), \label{13}
\eeq
\beq
w(x,y,t)=\sum_{i=1}^N (A_i(y,t)\zeta(x-x_i(y,t))+B_i(y,t)\wp(x-x_i(y,t))).
\label{13a}
\eeq
where $\rho_i$ is a rank-one matrix-function depending
on $y,t$
\beq
\rho_i=a_ib_i^+,\ \  {\rm i.e.}\ \ \
\rho_{i,\alpha}^{\beta}=a_{i,\alpha}b_i^{\beta}. \label{14}
\eeq

The precise relation is provided by the:

\begin{th}
Let us introduce the functions
\begin{eqnarray}
V(x)=\wp(x), \quad
\Phi(x,z)={\sigma(z-x)\over \sigma(z)\sigma(x)}e^{\zeta(z)x}.
\label{1.5}
\end{eqnarray}
The equations
\beq
\left(\p_t-\p_{x}^{2} + \sum_{i=1}^N a_i (t) b_i^+(t)V(x-x_i(t))
\right)\Psi=0 \label{1.3}
\eeq
\beq
\Psi^+ \left(\p_t -\p_x^2 + \sum_{i=1}^N a_i (t) b_i^+(t)V(x-x_i(t)) \right)=0
\label{1.3a}
\eeq
(where we define $\Psi^+ \p \equiv -\p \Psi^+$)
have solutions $\Psi, \Psi^+$ of the form
\beq
\Psi=\sum_{i=1}^N s_i(t,k,z)\Phi(x-x_i(t),z) e^{kx+k^2t}, \label{1.4}
\eeq
\beq
\Psi^+=\sum_{i=1}^N s_i^+(t,k,z)\Phi(-x+x_i(t),z) e^{-kx-k^2t}, \label{1.4a}
\eeq
where $s_i$ and $s_i^+$ are  $l$-dimensional vector $s_i=(s_{i,\alpha})$ and
co-vector $s_i^+=(s_i^{\alpha})$, respectively,
if and only if $x_i(t)$ satisfy the equations (\ref{4}) and  the vectors
$a_i,\ b_i^+$ satisfy the constraints (\ref{7}) and  the system
of equations
\beq
\dot{a}_i= -\sum_{j\neq i}a_j(b_j^+a_i)V(x_i-x_j)-\lambda_i a_i,\label{1.6}
\eeq
\beq
\dot{b}_i^+=\sum_{j\neq i}b_j^+(b_i^+a_j)V(x_i-x_j)+ \lambda_i
b_i^+\label{1.7}
\eeq
where $\lambda_i=\lambda_i(t)$ are scalar functions.
\end{th}

{\bf Remark 1.} The system (\ref{4},\ref{1.6},\ref{1.7}) is
``gauge equivalent" to the system (\ref{4}-\ref{6}). This means that
if $(x_i, a_i,  b_i^+)$ satisfy the equations
(\ref{4},\ref{1.6},\ref{1.7}) then $x_i$ and the vector-functions
\beq
\hat a_i=a_iq_i,
\ \hat b_i^+=b_iq_i^{-1},\ \ q_i=\exp(\int^t \lambda_i(t)dt) \label{1.10}
\eeq
are solutions of the system (\ref{4}-\ref{6}).

{\bf Remark 2.} In the scalar case the ansatz (\ref{1.3}) was introduced in
\cite{kr2}. Its particular form  was inspired by the well-known
formula  for the solution of the Lam\'e equation:
\beq
\bigl({d^2\over dx^2}-2\wp(x)\bigr)\Phi(x,z)=\wp(z)\Phi(x,z). \label{1.11}
\eeq

\noindent {\it Proof.}
Inserting equation~(\ref{1.4}) into equation~(\ref{1.3}) we find the
condition:
\begin{eqnarray*}
A&\equiv& \sum_{i=1}^N
\left\{  \vphantom{\sum_{j=1}^N}    \dot{s}_i \Phi(x-x_i,z) -(\dot{x}_i+2k)s_i
\Phi'(x-x_i,z)-s_i\Phi''(x-x_i,z)\right .\\
&+&\left . \sum_{j=1}^N a_j (b_j^+ s_i) \wp(x-x_j)\Phi(x-x_i,z)
\right\}=0
\end{eqnarray*}
where $\Phi'=\partial_x \Phi$ and so on.

The vanishing of the triple pole $(x-x_i)^{-3}$ gives the condition:
\begin{equation}
a_i (b_i^+ s_i)=2s_i \label{triple}
\end{equation}
Using this condition and the Lam\'e equation~(\ref{1.11}) we can
identify the double pole $(x-x_i)^{-2}$. Its vanishing gives the
condition:
\begin{equation}
s_i(\dot{x}_i+2k)+\sum_{j\neq i}a_ib_i^+ s_j\Phi(x_i-x_j,z)=0
\label{double}
\end{equation}
We finally identify the residue of the simple pole and obtain the
condition:
\beq
\dot{s}_i+\left(\sum_{j\neq i}a_jb_j^+\wp(x_i-x_j)-\wp(z)\right)s_i+
a_i\sum_{j\neq i}(b_i^+s_j)\Phi'(x_i-x_j,z)=0. \label{simple}
\eeq

Inserting now equations~(\ref{triple},\ref{double},\ref{simple}) into
the expression of $A$ one sees that $A$ vanishes identically due to
the functional equation:
\begin{equation}
\Phi'(x,z)\Phi(y,z)-\Phi(x,z)\Phi'(y,z)=(\wp(y)-\wp(x))\Phi(x+y,z)
\label{Calogero}
\end{equation}
We have shown that the function $\psi$ given by eq.~(\ref{1.4})
satisfies equation~(\ref{1.3}) if and only if the
conditions~(\ref{triple},\ref{double},\ref{simple}) are fulfilled.

Equation~(\ref{triple}) implies that the vector $s_i$ is
proportional to the vector $a_i$. Hence:
\begin{equation}
s_{i,\alpha}(t,k,z)=c_i(t,k,z)a_{i,\alpha}(t) \label{1.13}
\end{equation}
Moreover from~(\ref{triple}) it follows that the constraints~(\ref{7})
should be fulfilled.

Equation~(\ref{double}) can then be rewritten as a matrix equation for
the vector $C=(c_i)$:
\begin{equation}
(L(t,z)+2kI)C=0 \label{Llax}
\end{equation}
where the Lax matrix $L(t,z)$ with spectral parameter $z$ is given by:
\beq
L_{ij}(t,z)=\dot{x}_i\delta_{ij}+(1-\delta_{ij})f_{ij}\Phi(x_i-x_j,z).
\label{1.17}
\eeq

We can rewrite equation~(\ref{simple}) as:
\begin{equation}
\dot{a}_i=-\lambda_i a_i-\sum_{j\neq i} a_j(b_j^+a_i)\wp(x_i-x_j)
\label{eqsa}
\end{equation}
where we have defined:
$$\lambda_i={\dot{c}_i\over c_i}-\wp(z)+\sum_{j\neq i}(b_i^+a_j)
\Phi'(x_i-x_j,z) {c_j\over c_i}$$
But this last equation can be rewritten:
\begin{equation}
(\partial_t+M)C=0\label{Mlax}
\end{equation}
where the second element $M$ of the Lax pair is given by:
\beq
M_{ij}(t,z)=(-\lambda_i-\wp(z))\delta_{ij}+
(1-\delta_{ij})f_{ij}\Phi'(x_i-x_j,z). \label{1.18}
\eeq

The same arguments show that the existence of a solution $\Psi^+$ of the form
(\ref{1.4a}) implies (cancellation of the triple pole):
\beq
s_i^{\alpha}=c_i^+ b_i^{\alpha}, \label{1.13a}
\eeq
and the covector $C^+=(c_i^+)$ satisfies the equation (cancellation of
the double pole):
\beq
C^+(L(z)+2k)=0. \label{1.19a}
\eeq
Finally looking at the simple pole one gets:
\begin{equation}
\dot{b}_i^+=\lambda_i^+ b_i^++\sum_{j\neq i}
(b_i^+a_j)b_j^+\wp(x_i-x_j) \label{eqsb}
\end{equation}
with a new scalar $\lambda_i^+$ given by:
$$\lambda_i^+=-{\dot{c}_i^+\over c_i^+}-\wp(z)
+\sum_{j\neq i}{c_j^+\over c_i^+}(b_j^+a_i)\Phi'(x_j-x_i)$$
The equations~(\ref{eqsa},\ref{eqsb}) are compatible with
$f_{ii}=b_i^+a_i=2$ only when $\lambda_i^+=\lambda_i$. Finally we can
rewrite the definition of $\lambda_i^+$ as:
\beq
\partial_t C^+ -C^+M=0.\label{1.19b}
\eeq

To end the proof of the theorem we have to establish that the $x_i(t)$
satisfy equation~(\ref{4}). For this we exploit the compatibility
conditions between eq.~(\ref{Llax},\ref{Mlax}) and between
eq.~(\ref{1.19a},
\ref{1.19b}) which read respectively:
$$(\dot{L}+[M,L])C=0 \quad C^+(\dot{L}+[M,L])=0$$
Computing $\dot{L}+[M,L]$ we see that the off-diagonal elements vanish
identically due to equations~(\ref{1.6},\ref{1.7}) while the diagonal
elements are precisely the equations of motion of the $x_i$.
The computation uses again equation~(\ref{Calogero}) and we have
therefore shown the Lax form of the equations of motion:
\beq
\dot{L}=[L,M] \label{Laxform}
\eeq
\square

{\bf Remark.} In \cite{c} it was proved that Lax equation (\ref{Laxform})
with the matrices $L$ and $M$ given by the
formulae (\ref{1.17}, \ref{1.18}) (with $f_{ij}=2, \lambda_i=0$)
is equivalent to the equations of motion of
the Calogero-Moser system if and only if the
functional equation (\ref{Calogero}) is fulfilled. In
\cite{c} the particular solutions of the functional equation
corresponding to the values $z=\omega_l$ was found. The proof of
this equation for arbitrary values of the
spectral parameter $z$ was given in \cite{kr2}.

Let us comment on the trigonometric and rational limits of the above formulae.
The trigonometric limit is obtained when one of the periods
$\omega\to\infty$. We choose
the other one as $i\pi$. In this limit the function $\Phi$ becomes:
$$\Phi(x,z)=\left({\coth\, x}-{\coth\, z}\right)e^{x\coth\,z}$$
The exponential factor in $\Phi$ comes
from the factor $\exp(\zeta(z)x)$ in the elliptic case which is
necessary to induce the double periodicity of $\Phi$ in $z$. In the
trigonometric case however it can be absorbed into a redefinition of $k$
and $s_i$ of the form:
$$k\to k-{\coth\,z}\qquad s_i\to s_i \exp\,\left({x_i(t)\coth\,z}+
{2kt\coth\,z} -{t\coth^2 z}\right)$$
and similarly for the dual quantities. In the following we shall
therefore remove this exponential factor in the definition of $\Psi$.
The definitions of the functions $V(x)$ and $\Phi(x,z)$ become
\begin{eqnarray}
V(x)={1\over \sinh^2 (x)}, \quad
\Phi(x,z)= \coth x -\coth z,
\label{1.5trig}
\end{eqnarray}
With these new functions, the above theorem remains valid, but due to
the redefinition of the function $\Phi(x,z)$, the expression of the
Lax matrices is slightly modified and reads:

\beq
L_{ij}(t,z)=(\dot{x}_i-{2\coth\,z})\delta_{ij}+(1-\delta_{ij})f_{ij}
\Phi(x_i -x_j, z)\label{Ltrig}
\eeq
\beq
M_{ij}(t)=-\lambda_i\delta_{ij}-(1-\delta_{ij})f_{ij}
V(x_i -x_j)\label{Mtrig}
\eeq

The rational limit is obtained straightforwardly from the trigonometric limit
by sending the second period $\omega' \to\infty$.
The functions $V(x)$ and $\Phi(x,z)$ become
\begin{eqnarray}
V(x)={1\over x^2}, \quad
\Phi(x,z)= {1\over x}- {1\over z},
\label{1.5rat}
\end{eqnarray}
and of course $\coth\,z \to 1/z$ in eq~(\ref{Ltrig}).

Notice that as compared to the
elliptic case there is a decoupling between the spectral parameter
$z$ and the $x_i$'s in the Lax matrix~(\ref{Ltrig}).

\part{The Direct Problem}

\section{The spectral curve.}

Due to equation~(\ref{Llax}) the parameters $k$ and $z$ are
constrained to obey:
\beq
R(k,z)\equiv \det\,(2kI+L(t,z))=0 \label{spec}
\eeq
This defines a curve $\Gamma$ which is time--independent due to the
Lax equation~(\ref{Laxform}). This curve plays a fundamental role in
the subsequent analysis. Its properties are different in the elliptic,
trigonometric
and rational cases. Remark moreover that $\Gamma$ is invariant
under the symmetries (\ref{local},\ref{global}).

\begin{prop}
In the elliptic case we have:
\beq
R(k,z)=\sum_{i=0}^N r_i(z)k^i \label{1.25}
\eeq
where the $r_i(z)$ are elliptic functions of $z$, independent of $t$,
having the form:
\beq
r_i(z)=I_i^0+\sum_{s=0}^{N-i-2}I_{i,s}\p^s_z \wp(z). \label{1.260}
\eeq
In a neighbourhood of $z=0$ the function $R(k,z)$ can be
represented in the form:
\beq
R(k,z)=2^N\prod_{i=1}^N(k+\nu_i z^{-1}+h_i(z)), \label{1.29}
\eeq
where $h_i(z)$ are regular functions of $z$ and
\beq
\nu_i=1 , \ \ i>l.\label{1.30}
\eeq
\end{prop}

\noindent{\it Proof.} The matrix elements (\ref{1.17}) are
double periodic functions
of the variable $z$ having an essential singularity at $z=0$, but the functions
$r_i(z)$
are meromorphic because $L(t,z)$ can be represented in the form
\beq
L(t,z)=G(t,z)\tilde L(t,z) G^{-1}(t,z),
\ G_{ij}=\delta_{ij}\exp(\zeta(z)x_i(t))  ,\label{1.26}
\eeq
where $\tilde L_{ij}(t,z)$ are meromorphic functions of the variable $z$ in
a neighbourhood of the point $z=0$. In fact we have:
\beq
\tilde{L}(t,z)=-{1\over z}(F(t)-2I)+O(z^0)\label{resell}
\eeq
where $F(t)$ is the matrix of elements $f_{ij}(t)$.
Therefore $r_i(z)$ are elliptic
functions having poles of degree $N-i$ at most
at the point $z=0$.
Hence they can be represented in the form~(\ref{1.260})
as a linear combination of the function $\wp(z)$ and its derivatives.
The coefficients $I_i^0,I_{i,s}$  of this expansion are the integrals
of motion
of the system (\ref{1}).
Each set of given values of these integrals defines
an algebraic curve $\Gamma$.

Since around $z=0$ the function $r_i(z)$ has a pole of order $N-i$, a
factorization of the form~(\ref{1.29}) holds. Due to equation~(\ref{resell})
the coefficients $-2\nu_i$ in eq.(\ref{1.29}) are the eigenvalues
of the matrix $F-2I$. From
eq.~(\ref{8}) we see that $F$ is of rank $l$, hence the eigenvalue
$\nu_i=1$ has multiplicity $N-l$.
Moreover the corresponding $(N-l)$-dimensional subspace of
eigenvectors $C=(c_1,\ldots,c_N)$ is defined by the equations
\beq
\sum_{j=1}^N c_j a_{j,\alpha}=0 , \ \ \ \alpha=1,\ldots,l. \label{1.28}
\eeq
\square

{\bf Remark.}
The conditions~(\ref{1.30}) imply a full set of linear relations on the
integrals $I_i^0, I_{is}$ of the system (\ref{1}).
Let us take any polynomial (in $k$) $R(k,z)$ of the from
(\ref{1.25}) with $r_i(z)$ of the form (\ref{1.260}). It depends on
$N(N+1)/ 2$ parameters $I_i^0, I_{is}$.
Let us introduce the variable $\tilde k=k+z^{-1}$. Then the polynomial
in this variable $\tilde R(\tilde k,z)= R(\tilde k-z^{-1},z)$ for a
generic set of variables $I_i^0, I_{is}$ can be represented in the form
\beq
\tilde R(\tilde k,z)=\sum_{i=0}^N \tilde R_i(\tilde k)z^{-i} +
{\cal R}(z,\tilde k), \label{spain1}
\eeq
where $\tilde R_i$ are polynomials in $\tilde k$ of degree ${\rm deg}\, \tilde
R_i=N-i$ and ${\cal R}(z,\tilde k)=O(z)$ is a regular series in $z$ with
coefficients
that are polynomials in $\tilde k$ of degree $N-1$. The conditions (\ref{1.30})
imply that
\beq
\tilde R_i(\tilde k)=0, \ \ i>l. \label{spain2}
\eeq
The coefficients of $\tilde R_i$ are linear combinations of the parameters
$I_i^0, I_{is}$. Therefore, (\ref{spain2}) is equivalent to a set of
$(N-l)(N-l+1)/2$ linear equations on these parameters. The total number
of independent parameters is therefore equal to $Nl-l(l-1)/2$ which is exactly
half
the dimension of the reduced phase space.

In the trigonometric and rational cases the parametrization of the
corresponding
spectral curve is even more explicit.

\begin{prop}
In the trigonometric case we have:
\beq
R(k,z)=R_0(k)+{\coth\,z}R_1(k)+\cdots+{\coth^l z}R_l(k) \label{gammatrig}
\eeq
where the $R_m(k)$ are polynomials in $k$ of degree
deg$_k\,R_m = N-m$ and
\beq
R(k,z=-\infty)=R(k+2,z=+\infty)\label{reltrig}
\eeq
In a neighbourhood of $z=0$ the function $R(k,z)$ can be
factorized in the form of eq.(\ref{1.29}) where now
$\nu_i=0 , \ \ i>l$.
\end{prop}

\noindent{\it Proof.} The matrix $L(t,z)$ depends on $z$ only through the
term $\coth z F$. Since $F$ is of rank $l$,  $R(k,z)$ is of the
form~(\ref{gammatrig}). To prove the relation~(\ref{reltrig}) it is enough to
remark
that:
\beq
L(t,-\infty)+2kI=e^{2X}\left(L(t,+\infty)+2(k+2)I\right)e^{-2X}\label{relL}
\eeq
with $X={\rm Diag}\,(x_i(t))$.
The conditions $\nu_i =0, i> l$ follow from the fact that around $z=0$
we now have:
\beq
L(t,z)= -{1\over z}F + O(z^0) \label{residutrig}
\eeq
\square

\begin{prop}
In the rational case we have:
\beq
R(k,z)=R_0(k)+{1\over z}R_1(k)+\cdots+{1\over z^l}R_l(k) \label{gammarat}
\eeq
where the $R_m(k)$ are polynomials in $k$ of degree
deg$_k\,R_m = N-m$ and
\beq
R_1(k)=-{ {\rm d}R_0(k)\over {\rm d}k} \label{relrat}
\eeq
In a neighbourhood of $z=0$ the function $R(k,z)$ can be
factorized in the form of eq.(\ref{1.29}) where now
$\nu_i=0 , \ \ i>l$.
\end{prop}

\noindent{\it Proof.}
Since
\begin{eqnarray}
[X,L(t,\infty)]=F-2I
\label{XL}
\end{eqnarray}
we have:
\begin{eqnarray}
L(t,z)+2kI&=&L(t,\infty)-{1\over z}[X,L(t,\infty)]+2(k-{1\over z})I \\
&=&(I-{1\over z}X)(L(t,\infty)+2(k -{1\over z})I)(I-{1\over z}X)^{-1}
+O({1\over z^2})
\end{eqnarray}
hence $R(k,z)=R_0(k-{1\over z})+O({1\over z^2})$ so that
$R_1=-R_0'$. \square

As a consequence we can count the number of parameters entering the
spectral curve. Each $R_m$ depends on $N-m+1$ parameters, but
relations~(\ref{reltrig}) or (\ref{relrat}) remove $N$ parameters
and the leading term of $R_0$ is already given so that we get $Nl
-l(l-1)/2$ parameters, which is exactly half the dimension
of the reduced phase space. These parameters can be identified with the
action variables of our model and are in involution since there exists
an $r$--matrix for $L$~\cite{AvBaBi}.

We now compute the genus of the spectral curve $\Gamma$.

\begin{prop}
For generic values of the action variables the genus of the spectral
curve is given by:
\begin{eqnarray}
 Elliptic~ case:\quad
g&=&Nl-{l(l+1)\over 2}+1 \label{genell} \\
  \matrix{ Trigonometric \cr
	   and~ rational~ cases}: \quad
g&=&N(l-1)-{l(l+1)\over 2}+1 \label{genrat}
\end{eqnarray}
\end{prop}

\noindent{\it Proof.}
Equation~(\ref{spec})
allows to present the compact Riemann surface $\Gamma$ as an $N$-sheeted
branched covering of the base curve of the variable
$z$, i.e., the completed plane in the trigonometric and rational cases and the
torus in
the elliptic case.
The sheets are the $N$ roots in $k$. By the
Riemann--Hurwitz formula we have $2g-2=N(2g_0-2)+\nu$ where
$g_0$ is the genus of the base curve, i.e. $g_0=0$ in the trigonometric and
rational
cases, $g_0=1$ in the elliptic case.
Here $\nu$ is the
number of branch points, i.e. the number of values of $z$ for
which $R(k,z)$ has a double root in $k$. This is the number of
zeroes of $\partial_k R(k,z)$ on the surface $R(k,z)=0$.
But $\partial_k R(k,z)$ is a meromorphic function on the
surface, hence it has as many
zeroes as poles. The poles are located above $z=0$ or
$k=\infty$ which is the same, and are easy to count.

Let $P_i$ be the points of $\G$ lying on the different sheets over the
point $z=0$. In the neighbourhood of $P_i$
 the function $k$ has
the expansion
\beq
k_i=-\nu_i z^{-1}-h_i(z).\label{1.31}
\eeq
Hence, the function $\p R /\p k$ in the neighbourhood of $P_i$ has the form
\beq
\p R /\p k=2^N \prod_{j\neq i}((\nu_j-\nu_i)z^{-1}+(h_j(z)-h_i(z))).
\label{1.31a}
\eeq
 From this we see that on
each of the $l$ sheets $(k_i(z),z)$  ($i=1,\cdots,l)$ we have
one pole of order $(N-1)$.
On each of the $(N-l)$ sheets   $(k_i(z),z)$ ($i=l+1,\cdots,N$) we have
one pole of order $l$. Finally $\nu=l(N-1)+(N-l)l$ in either case.
Inserting this value in the Riemann--Hurwitz theorem yields the
result.
\square

\section{Analytic properties of the eigenvectors of the Lax matrix.}

For a generic point $P$ of the curve $\G$,
i.e. for the pair $(k,z)=P$, which satisfies
the equation (\ref{1.25}), there exists at time $t=0$
a unique eigenvector $C(0,P)$ of the
matrix $L(0,z)$ normalized by the condition $c_1(0,P)=1$.
In fact the un--normalized components $c_i(0,P)$
can be taken as $\Delta_i (0,P)$
where $\Delta_i (0,P)$ are suitable minors of the matrix $L(0,z)+2k I$,
and are thus
holomorphic functions on $\Gamma$ outside the points above $z=0$. After
normalizing the first component, all the other
coordinates $c_j(0,P)$ are meromorphic functions on $\G$, outside
the points $P_i$ above $z=0$. The poles of $c_j(0,P)$ are the zeroes on
$\G$ of the first minor of the matrix $L(0,z)+2kI$, i.e.,
they are defined by the system of the equation (\ref{1.25}) and the equation
\beq
\det(2k\delta_{ij}+L_{ij}(0,z))=0,\ \ i,j>1. \label{1.321}
\eeq
Thus the position of these poles only depend on the initial data.

In the trigonometric and
rational cases nothing particular
happens above $z=0$. In the elliptic case however one has to be careful
because of the essential singularity.

\begin{prop}
In the elliptic case, in the
neighbourhood of the point
$P_i$ the coordinate $c_j(0,P)$ has the form
\beq
c_j(0,P)=(c_j^{(i)}(0)+O(z))\exp\,[\zeta(z)(x_j(0)-x_1(0))],\label{1.33}
\eeq
where
$c_j^{(i)}(t)$ is the eigenvector of
the matrix $F(t)$ corresponding to the non--zero
eigenvalue $2(1-\nu_i)$ i.e.,
\beq
\sum_{j=1}^N f_{kj}(t)c_j^{(i)}(t)=2(1-\nu_i) c_k^{(i)}(t) .\label{1.331}
\eeq
\end{prop}

\noindent {\it Proof.} From equation~(\ref{1.26}), we have
$C(0,P)= G(0,z) \tilde C(0,P)$, where $\tilde C(0,P)$ is an eigenvector
of $\tilde L(0,z)$. Using equation~(\ref{resell}), we have
$ \tilde C (0,P)= \tilde C^{(i)} + O(z)$ where $\tilde C^{(i)}$ is an
eigenvector of $(F-2I)$. Therefore we have
$c_j(0,P)=(c_j^{(i)}(0)+O(z))\exp{(\zeta(z)x_j(0))}$. Normalizing $c_1(0,P)=1$
yields the result.  \square

We can now compute the number of poles of $C$ on $\G$. This number is
the same in all cases, although its relation to
the genus of $\Gamma$ differs in the elliptic and other cases.

\begin{prop}
The number of poles of $C(0,P)$ is:
\begin{eqnarray}
m&=&Nl-{l(l+1)\over 2}=g  - 1 ~~~~~~~
\quad{\rm Elliptic ~case}\label{mell}\\
m&=&Nl-{l(l+1)\over 2}=g+N-1 \quad{\rm Trigonometric~and~rational
{}~cases}
\label{mrat}
\end{eqnarray}
\end{prop}

\noindent{\it Proof.}
Let us introduce the function $W$ { of the
complex variable} $z$ defined by:
$$W(z)=\left( {\rm Det}\left| c_i(M_j) \right| \right)^2$$
where the $M_j$'s are the $N$ points above $z$. It is
well--defined on the base curve
since the Det$^2$ does not depend on the order of the
$M_j$'s.

In the trigonometric and rational cases it is a meromorphic function,
hence has the same number of zeroes and poles. In the elliptic case it
has an essential singularity at $z=0$ of the form
$\exp\,2\zeta(z)\sum (x_i(0)-x_1(0))$. This does not affect the
property that the number of poles is equal to the number of zeroes.
Clearly $W$ has a double pole for values of $z$ such
that there exists above $z$ a point $M$ at which $C(M)$ has
a simple pole.

We show that
$W(z)$  has a simple zero for values of $z$ corresponding to
a branch--point of the covering, hence $m=\nu/2$.

First notice that $W(z)$ only vanishes on branch--points, where
there are at least two identical columns. Indeed, let
$M_i=(k_i,z)$ be the $N$ points above $z$. Then the $C(M_i)$ are
the eigenvectors of
$L(z)$ corresponding to the eigenvalues $-2k_i$ hence are
linearly independent when all the $k_i$'s are different.
Therefore $W(z)$ cannot vanish at such a point.
Let us assume now that $z$
corresponds to a branch point, which is generically of order $2$.
At such a point $W(z)$ has a simple zero. Indeed let
$\xi$ be an analytical parameter on the curve around the branch point.
The covering projection $M\to z$ gets expressed as
$z=z_0+z_1 \xi^2 +O(\xi^3)$. The determinant vanishes to
order $\xi^1$ hence $W$ vanishes to order $\xi^2$, but this is precisely
proportional to $z -z_0$. \square

At this point the analysis of the elliptic and the trigonometric
and rational cases
begin to differ substantially. We treat them separately.

\subsection{The elliptic case}

In this case we compute the time evolution of the above eigenvectors.
\begin{prop}
The coordinates $c_j(t,P)$ of the vector-function $C(t,P)$ are
meromorphic functions on $\G$ except at the points $P_i$. Their
poles $\gamma_1,\ldots,\gamma_{g-1}$ do not depend on $t$. In
the neighbourhood of $P_i$ they have the form
\beq
c_j(t,P)=c^{(i)}_j(t,z)\exp{(\zeta(z)(x_j(t)-x_1(0))+\mu_i(z)t)},\label{1.37}
\eeq
where $c^{(i)}_j(t,z) $ are regular functions of $z$
\beq
c^{(i)}_j(t,z) =c_j^{(i)}(t)+O(z)\label{1.38}
\eeq
and
\beq
\mu_i(z)=(1-2\nu_i)z^{-2}-2h_i(0)z^{-1}+O(z^0). \label{1.381}
\eeq
\end{prop}
\noindent{\it Proof.}
The fundamental matrix $S(t,z)$ of solutions to equation
\beq
(\p_t+M(t,z)) S(t,z)=0, \ \  S(0,z)=1, \label{1.34}
\eeq
is a holomorphic function of the variable $z$ for $z\neq 0$. At
$z=0$ however it has an essential singularity.

We have $L(t,z)= S(t,z) L(0,z) S^{-1}(t,z)$. Therefore the
vector $C(t,z)= S(t,z)C(0,z)$ is
the common solution to (\ref{Mlax}) and to the equation
\beq
(L(t,z)+2kI)C(t,P)=0,\ \ P=(k,z)\in \G. \label{1.36}
\eeq
Since $S(t,z)$ is regular for $z\neq 0$ we see that $C(t,P)$ has the
same poles as $C(0,P)$.

Let us consider the vector $\tilde C(t,P)$ defined as
\beq
C(t,P)=G(t,z)\tilde C(t,P), \label{1.382}
\eeq
where $G(t,z)$ is the same as in (\ref{1.26}). This vector is an eigenvector of
the matrix $\tilde L(t,z)$ and satisfies the equation
\beq
(\p_t+\tilde M(t,z))\tilde C(t,P)=0,\ \
\tilde M=G^{-1}\p_tG+G^{-1}MG. \label{1.383}
\eeq
 From (\ref{1.17},\ref{1.18}) it follows that
\beq
\tilde M(t,z)=-z^{-2}I +z^{-1}\tilde L(t,z)+O(z^0). \label{1.384}
\eeq
It follows from this relation that around $P_i$
$$\partial_t \tilde C (t,z) = (\tilde \mu_i (t,z) +O(z^0))
\tilde C (t,P) $$
where
\beq
\tilde \mu_i(t,z)=
z^{-2}+2k_i(z)z^{-1}=(1-2\nu_i)z^{-2}-2h_i(0)z^{-1}+O(z^0).
\label{1.385}
\eeq
 From this, we deduce that around $P_i$, we have
\begin{eqnarray}
\tilde C (t,P) = e^{ \mu_i (z)t} \hat C (t,P)
\nonumber
\end{eqnarray}
where the vector $\hat C (t,P)$ is regular around $P_i$. Multiplying
by $G(t,z)$ and normalizing $c_1(0,P)=1$, we get the result.
\square

\subsection{The trigonometric and rational cases}

In these cases $M$ is constant on the curve and we can choose:
\begin{equation}
C(t,P)=S(t)C(0,P) \label{gpt}.
\end{equation}
where $S(t)$ is defined as above and is independent of the point of the curve.
Hence $C(t,P)$ is a meromorphic vector with the same poles as
$C(0,P)$. Moreover since $C(0,P)$ is regular above $z=0$ the same is
true for $C(t,P)$.

However, there appear new points at infinity playing the major
role.

In the trigonometric case, we have two series of such points
above $z=\pm\infty$. Let us denote these points by $Q_j(k=\chi_j,z=-\infty),\;
j=1,\cdots,N$ and $T_j(k=\chi_j+2,z=+\infty),\;j=1,\cdots,N$, (the
$k$-coordinate of $T_j$ is $\chi_j +2$ because of equation~(\ref{reltrig})).

In the rational case, we have only one series of such points
above $z=\infty$. We denote them by
$Q_j(k=\chi_j, z=\infty),\; j=1,\cdots,N$.

In the trigonometric case the base curve is in fact a cylinder, i.e.,
a sphere with two marked points, while in the rational case
it is a sphere with only one marked point.

We study the
solutions of the equation $$L(t,P)C(t,P)\equiv
\left(L(t,z)+2kI\right)C(t,P)=0$$
around these points.

\begin{prop}
In the trigonometric case, the eigenvectors at the points $Q_j$ and $T_j$
are related by:
\beq
C(t,T_j)=\mu_j e^{-4(\chi_j +1)t}e^{-2X}C(t,Q_j)\label{relC}
\eeq
In the rational case, at the point $Q_j$, we have:
\begin{eqnarray}
\partial_k C(t,Q_j) = -(X+2 \chi_j t - \mu_j)  C(t,Q_j).
\label{constr}
\end{eqnarray}
The parameters $\mu_j$ are constants and $X ={\rm Diag~}(x_i(t))$.
Moreover with the normalization $c_1(0,P)=1$ all the $\mu_j$'s are
equal to $e^{2x_1(0)}$ or $x_1(0)$ respectively.
\end{prop}

\noindent{\it Proof.} Let us prove first eq.(\ref{relC}).
 From equation~(\ref{relL}) we see that:
$$C(t,T_j)=\mu_j(t)e^{-2X}C(t,Q_j)$$
To compute $\mu_j(t)$ we exploit the Lax equation $\dot{C}=-MC$ at the
points $Q_j$, $T_j$ using the fact that $M$ is independent of the point
on $\G$. Using the relation:
$$e^{-2X}M(t)e^{2X}=M(t)+2L(t,+\infty)-2\dot{X}+4I$$
we find $\dot{\mu}_j=-(4\chi_j+4)\mu_j$.

The proof of eq.(\ref{constr}) is slightly more complicated.
First of all, around  a point $Q_j$, the curve has the equation:
$$R_0(k)-{1\over z}R_0'(k)+O({1\over z^2})=0$$
implying
\begin{equation}
\left.{1\over z^2}{d z\over dk}\right|_{Q_j}=-1 \Longrightarrow
{1\over  z} = (k -\chi_j)+ O(k -\chi_j)^2
\label{dkalpha}
\end{equation}
hence $k$ is an analytic parameter around $Q_j$.
%In particular
%\begin{eqnarray}
%{1\over  z} &=& (k -\chi_j)+ O(k -\chi_j)^2 \label{1suralpha}\\
%C(t,k, z)&=& C(t,Q_j) +(k-\chi_j) \partial_k C(t,Q_j) +
%O(k-\chi_j)^2. \nonumber
%\end{eqnarray}

Next we consider the equation:
$
\relax \left[L(t,\infty) +2kI -{1\over  z} F \right] C(t,P)=0.
$
It gives:
\begin{eqnarray}
%L(t,Q_j)  C(t,Q_j) &=&0;\label{Lr} \\
L(t,Q_j)  \partial_k C(t,Q_j)&=& (F-2I) C(t,Q_j).
\label{Ls}
\end{eqnarray}
To solve this equation, we remark that by virtue of
equation (\ref{XL}),
we have
$$
L(t,Q_j)( -X  C(t,Q_j))
=[X,L(t,Q_j)]  C(t,Q_j)= (F-2I) C(t,Q_j)
$$
therefore the general solution of eq.~(\ref{Ls}) is of the form
\begin{eqnarray}
\partial_k C(t,Q_j) = -X  C(t,Q_j) + \mu_j(t)  C(t,Q_j).
\label{s}
\end{eqnarray}
To find the functions $\mu_j(t)$, we use the evolution equation
$\dot{C}= -M {C}$, which implies:
$$
\dot{\mu}_j C(t,Q_j) = (\dot{X}- [X,M]) C(t,Q_j)
$$
but it is straightforward to check that $\dot{X}- [X,M]= L(t,\infty)$.
Therefore
 $\dot{\mu_j} = -2 \chi_j$.
Applying equation~(\ref{constr}) for $i=1$ at $t=0$ we have $c_1=1$,
$\partial_k c_1=0$ hence all the $\mu_j$'s are equal to $x_1(0)$.\square

Similarly for the covector $C^+$ we have:
\begin{prop}
At the point $Q_j$, we have in the trigonometric case:
\begin{eqnarray}
C^+(t,T_j)=\mu_j^+ e^{4(\chi_j+1)t}C^+(t,Q_j)e^{2X}
\label{constrig}
\end{eqnarray}
and in the rational case we have:
\begin{eqnarray}
\partial_k C^+(t,Q_j) =  C^+(t,Q_j)(X +2 \chi_j t + \mu^+_j) .
\label{constra}
\end{eqnarray}
Moreover with the normalization $c^+_1(0,P)=1$ all the $\mu^+_j$'s are
equal to $e^{-2x_1(0)}$, or $-x_1(0)$ in the rational case.
\end{prop}

\section{The analytic properties of $\Psi$ and $\Psi^+$.}

In this section we encode the previous results on the eigenvectors
of the Lax matrix into analyticity properties of $\Psi$ and $\Psi^+$.
We treat separately the three cases.

\subsection{The elliptic case.}

\begin{th} The components $\Psi_{\alpha}(x,t,P)$ of the
solution $\Psi(x,t,P)$ to the nonstationary matrix  Schr\"odinger equation
(\ref{1.3}) are defined on the $N$-sheeted covering $\G$ of the initial
elliptic
curve. They are meromorphic on  $\G$ outside $l$ points
$P_i, \ i=1,\ldots,l$.
For general initial conditions the curve $\G$ is smooth, its genus equals
$g=Nl-{l(l+1)\over 2}+1$ and the $\Psi_{\alpha}$ have $(g-1)$
poles $\gamma_1,\ldots,\gamma_{g-1}$ which do not
depend on the variables $x,t$.
In a neighbourhood of $P_i, \ i=1,\ldots,l$, the function $\Psi_{\alpha}$
has the form:
\beq
\Psi_{\alpha}(x,t,P)=
(\chi_0^{\alpha i}+\sum_{s=1}^{\infty}\chi_s^{\alpha i}(x,t)z^{s})
e^{\lambda_i(z)x +\lambda_i^2(z)t}~\Psi_1(0,0,P),  \label{1.39}
\eeq
where
\beq
\lambda_i(z)=z^{-1}+k_i(z)=(1-\nu_i)z^{-1}-h_i(0)+O(z) \label{1.40}
\eeq
and $\chi_0^{\alpha i}$ are constants independent of t.
\end{th}

\noindent{\it Proof.}
We recall the relation between the function $\Psi$ and the
eigenvectors of the Lax matrix.
$$\Psi(x,t,P)=\sum_{j=1}^N s_j(t,P)\Phi(x-x_j(t),z) e^{kx+k^2t}, \ \
s_j(t,P)=c_j(t,P)a_j(t)$$
It is obvious that the $(g-1)$ poles $\gamma_k$ of the $c_i$'s are
time--independent poles of $\Psi$. To study the behaviour of $\Psi$
above $z=0$ we use the expansion of $\Phi$ at $z=0$:
\beq
\Phi(x,z)=\left(-{1\over z}+\zeta(x)+O(z)\right)e^{ \zeta(z) x}\label{dphiz}
\eeq
and the expansion of the eigenvectors, equation~(\ref{1.37}). We get:
\beq
\Psi_\alpha=\sum_{j=1}^N\left(-{1\over z}+\zeta(x-x_j(t))
+O(z)\right)a_{j,\alpha}(t)
c_j^{(i)}(t,z)e^{\lambda_i(z)x +\lambda_i^2(z)t}e^{- z^{-1} x_1(0)}.
\label{star}
\eeq
On the $(N-l)$ branches $i>l$ we see that $\lambda_i(z)$ is regular at
$z=0$ due to eq.~(\ref{1.30}) so that $\Psi$ has no essential
singularity at the $P_i$ for $i>l$ apart from the irrelevant constant factor
$\exp\,(- z^{-1} x_1(0))$.
%(one can get rid of it by dividing $\Psi$ by $\Phi(x_1(0),z)$)
Even more there is no pole at these
points. This is because $\sum_j a_j^\alpha(t)c_j^{(i)}(t,z)=O(z)$
due to equation~(\ref{1.28}).

It only remains to prove that the leading term of the
expansion of the first factor in the right hand side of (\ref{1.39}) does not
depend on $t$. (It does not depend on $x$ because the singular part of
$\Phi$ at $z=0$ does not depend on $x$.)
The substitution of the right hand side
of (\ref{1.39}) with $\chi_0^{\alpha j}=\chi_0^{\alpha j}(t)$ into the equation
\beq
(\p_t-\p_x^2+u(x,t))\Psi(x,t,P)=0, \label{1.41}
\eeq
gives that
\beq
u=2(\p_x\chi_1)\Lambda \chi_0^{-1}-(\p_t \chi_0)\chi_0^{-1},
\label{1.42}
\eeq
where  $\chi_s$ is a matrix with entries $\chi_s^{\alpha j}$ and $\Lambda$
is a diagonal matrix $\Lambda^{\alpha j}=(1-\nu_j)\delta^{\alpha j}$.
 From (\ref{star}) it follows that $\chi_1$ has the form
\beq
\chi_1=\sum_{i=1}^N R_i(t)\zeta(x-x_i(t)). \label{1.43}
\eeq
Therefore, for a potential of the form $u=\sum \rho_i(t)\wp(x-x_i(t))$, the
equality (\ref{1.43}) implies that
\beq
\rho_i=-2R_i(t)\Lambda \chi_0^{-1},\ \ \ \  (\p_t \chi_0)\chi_0^{-1}=0.
\label{1.44}
\eeq
Dividing eq.(\ref{star}) by the normalization factor $\Psi_1(0,0,P)$,
which plays no role in the fact that $\Psi(x,t,P)$ satisfies the
differential equation~(\ref{1.3}) we get the final result.
One can express $\chi_0$ in terms of the $c_i^{(j)}(t)$ defined in
eq.(\ref{1.331})
\beq
\chi_0^{\alpha j}=\sum_{i=1}^l c_i^{(j)}(0)a_{i,\alpha}(0) \label{1.400}
\eeq
\square

The same arguments show that:
\begin{th} The components $\Psi^{+,\alpha}(x,t,P)$ of the
solution $\Psi^+(x,t,P)$ to the nonstationary matrix  Schr\"odinger equation
(\ref{1.3a}) are defined on the same curve $\G$.
They are meromorphic on  $\G$
outside the $l$ punctures $P_i,\ i=1,\ldots,l$.
In general  $\Psi^{+,\alpha}$ have $(g-1)$
poles $\gamma^+_1,\ldots,\gamma^+_{g-1}$ which do not depend on the variables
$x,t$. In a neighbourhood of $P_i, \ i=1,\ldots,l$,
the function $\Psi^{+,\alpha}$ has the form:
\beq
\Psi^{+,\alpha}(x,t,P)=
(\chi_0^{+,\alpha i} +\sum_{s=1}^{\infty}\chi_s^{+,\alpha i}(x,t)z^{s})
e^{-\lambda_i(z)x -\lambda_i^2(z)t}~\Psi^{+,1}(0,0,P) ,  \label{1.39a}
\eeq
where the $\chi_0^{+,\alpha j}$ are constants.
\end{th}

{\bf Remark.}
Theorem 5.1 states, in particular, that the solution
$\Psi$ of equation (\ref{1.3}) is (up to normalization ) a Baker-Akhiezer
vector-function (\cite{kr3}). In the next section we show that this
function is uniquely defined by the curve
$\G$, its poles $\gamma_s$, the matrix $\chi_0$ and the value
$x_1(0)$.
All these values are defined by the initial Cauchy data and do not depend on
$t$. At the same time it is absolutely necessary to emphasize that part of them
depend on the choice of the normalization point $t_0$
that we choose as $t_0=0$.
Let us be more accurate. Any point of the phase space
$\{x_i,p_i,a_i,b_i^+| (b_i^+,a_i)=2\}$ defines the matrix $L$ with the help of
formulae (\ref{1.17}). The characteristic equation (\ref{1.25}) defines an
algebraic curve $\G$. The equation~(\ref{1.321}) defines a set
of $g-1$ points $\gamma_s$
on $\G$. Therefore, we may define a map
\beq
\{x_i,p_i,a_i,b_i^+| (b_i^+,a_i)=2\}\longmapsto \{ \G,\ \ D\in J(\G) \},
\label{1.50}
\eeq
\beq
D=\sum_{s=1}^{g-1}A(\gamma_s)+x_1U^{(1)} .\label{1.51}
\eeq
where $A:\G\to J(\G)$ is an Abel map and $U^{(1)}$ is a vector depending on
$\G$, only (see (\ref{2.25})).
The coefficients of the equation (\ref{1.25}) are integrals of the
hamiltonian system (\ref{1}). As we shall see in the next section
the second part of data (\ref{1.50}) define angle-type variables, i.e.
the vector
D(t) evolves linearly $D(t)=D(t_0)+(t-t_0)U^{(2)}$ if a point in phase space
evolves according to the equations (\ref{4}--\ref{6}). These equations
have the obvious symmetries:
\beq
a_i,b_i^+\to \lambda_i a_i,\lambda_i^{-1}b_i^+,\ \ a_i,b_i^+\to
W^{-1}a_i,b_i^+W,
\label{1.52}
\eeq
where $q_i$ are constants and $W$ is an arbitrary constant matrix.
In the next section we prove that to the data $\G,D$ one can associate
a unique point in the phase
space reduced under the symmetry~(\ref{1.52}).

\subsection{The trigonometric and rational cases.}

\begin{th} The components $\Psi_{\alpha}(x,t,P)$ of the
solution $\Psi(x,t,P)$ to the nonstationary matrix  Schr\"odinger equation
(\ref{1.3}) are defined on an $N$-sheeted covering $\G$ of the
completed complex plane.
They are meromorphic on  $\G$ outside $l$ points
$P_i, \ i=1,\ldots,l$.
For general initial conditions the curve $\G$ is smooth, its genus equals
$g=N(l-1)-{l(l+1)\over 2}+1$ and the $\Psi_{\alpha}$ have $(g+N-1)$
poles $\gamma_1,\ldots,\gamma_{g+N-1}$ which do not
depend on the variables $x,t$.
In a neighbourhood of $P_i, \ i=1,\ldots,l$, the function $\Psi_{\alpha}$
has the form:
\beq
\Psi_{\alpha}(x,t,P)=
(\chi_0^{\alpha i}z^{-1}+\sum_{s=1}^{\infty}\chi_s^{\alpha i}(x,t)z^{s-1})
e^{k_i(z)x +k_i^2(z)t},  \label{psirat}
\eeq
where
\beq
k_i(z)=-\nu_i z^{-1}-h_i(0)+O(z) \label{kiz}
\eeq
and $\chi_0^{\alpha i}$ are constants.

In the trigonometric case,  at the points
$T_j$, and $Q_j$ above $z=\pm\infty$ we have:
\begin{equation}
 \Psi(x,t, T_j)=  \mu_j \Psi(x,t,Q_j)\label{babeltrig}
\end{equation}

In the rational case, at the points $Q_j$ above $z=\infty$ we have:
\begin{equation}
 \partial_k \Psi (x,t,Q_j) = \mu_j \Psi (x,t,Q_j) \label{babel}
\end{equation}
where the $\mu_j$'s are defined in eqs.(\ref{relC},\ref{constr}).
\end{th}
\noindent{\it Proof.}
In the trigonometric case, the result follows
immediatly from equation~(\ref{relC}).

In the rational case, we use eq.(\ref{1.4}) with
$s_i(t,P) = a_i(t) c_i(t,P)$, we get:
$$\partial_k \Psi=\sum_i a_i\left\{
-{1\over z}c_i x - {1\over z}\partial_k
c_i+(1+{1\over z^2}\partial_k z -{2kt\over z})c_i
+{\partial_kc_i+(x_i+2kt)c_i\over x-x_i}\right\}e^{kx+k^2t}.$$
When $z=\infty$ we have $1+{1\over z^2}\partial_k z=0$
and:
$$\left . \partial_k \Psi\right |_{Q_j}=\left .
\left(\sum_i a_i {\partial_kc_i+(x_i+2kt)c_i\over x-x_i}
\right)e^{kx+k^2t}\right |_{Q_j}.$$
The result follows from equation~(\ref{constr}).\square

Similarly we have for $\Psi^+$:
\begin{th} The components $\Psi^{+,\alpha}(x,t,P)$ of the
solution $\Psi^+(x,t,P)$ to the nonstationary matrix  Schr\"odinger equation
(\ref{1.3a}) are defined on an $N$-sheeted covering $\G$ of the
completed complex plane.
They are meromorphic on  $\G$ outside $l$ points
$P_i, \ i=1,\ldots,l$ and  have $(g+N-1)$
poles $\gamma^+_1,\ldots,\gamma^+_{g+N-1}$ which do not
depend on the variables $x,t$.
In a neighbourhood of $P_i, \ i=1,\ldots,l$, the function $\Psi^{+,\alpha}$
has the form:
\beq
\Psi^{+,\alpha}(x,t,P)=
(\chi_0^{+,\alpha i}z^{-1}+\sum_{s=1}^{\infty}\chi_s^{+,\alpha i}(x,t)z^{s-1})
e^{-k_i(z)x -k_i^2(z)t},  \label{psirat+}
\eeq
and $\chi_0^{+,\alpha i}$ are constants. In addition at the points $T_j$,
$Q_j$ above $z=\pm \infty$ we have in the trigonometric case:
\beq
\Psi^+(T_j)=\mu_j^+\Psi^+(Q_j)
\eeq
and in the rational case we have at the points $Q_j$ above $z=\infty$:
\begin{equation}
\partial_k \Psi^+(x,t,Q_j)= \mu^+_j \Psi^+(x,t,Q_j) \label{babela}
\end{equation}
where the $\mu^+_j$'s are defined in equation~(\ref{constrig},\ref{constra}).
\end{th}

{\bf Remark.}
Obviously one can multiply the functions $\Psi_\alpha$ by a
meromorphic function $f(P)$ on $\Gamma$
without affecting the Schr\"odinger equation~(\ref{1.3}). We have already
used this property in equation~(\ref{1.39}) to factor out $\Psi_1(0,0,P)$. The
$m+l$ poles of the resulting Baker--Akhiezer function are now in arbitrary
position. In the trigonometric and rational cases we can use the same
feature in order to
match the normalizations used in the elliptic case. Let us define $f(P)$
with $m+l$ zeroes at the points $\gamma_1,\cdots, \gamma_m$ and $P_1,
\cdots, P_l$, $N$ poles at the points $Q_1,\cdots,Q_N$ and $l-1$ poles
$\gamma'_1 ,\cdots,\gamma'_{l-1}$ at some arbitrary prescribed
positions. This defines a divisor of degree $g$, and such a function
$f$ is uniquely determined. It has $g$ extra poles $\gamma'_l,\cdots,
\gamma'_{g+l-1}$. The function
\beq
\psi'=f\psi \label{funf}
\eeq
has now $m+l$
poles at the points $Q_j$ and $\gamma'_k$, and satisfies the
differential equation~(\ref{1.3}).
In the following sections we will use $\psi'$ (denoted $\psi$) and
$\gamma'_k$ (denoted $\gamma_k$).

\part{The inverse Problem}

\section{The elliptic case}

\subsection{The Baker--Akhiezer functions.}

At the begining of this section we present the necessary information
on the finite-gap theory \cite{kr3}.

\begin{th}
Let $\Gamma$ be a smooth algebraic curve of genus $g$ with fixed
local coordinates $w_i(P)$ in neighbourhoods of $l$ punctures
$P_i, \ w_i(P_i)=0, \ i=1,\ldots,l$. Then for each set of $g+l-1$ points
$\gamma_1,\ldots,\gamma_{g+l-1}$ in a general position there exists
a unique function $\psi_{\alpha}(x,t,P)$ such that

$1^0.$ The function $\psi_{\alpha}$ of the variable
$P\in \Gamma$ is meromorphic outside the punctures and has at most simple poles
at points
$\gamma_s$ (if all of them are distinct);

$2^0.$ In the neighbourhood of the puncture $P_j$ it  has the form
\beq \psi_{\alpha}(x,t,P)=e^{w_j^{-1}x+w_j^{-2}t}\left(\delta_{\alpha j}+
\sum_{s=1}^{\infty}\xi_s^{\alpha j}(x,t)w_j^s \right),\ \ \
w_j=w_j(P).\label{2.1}
\eeq
\end{th}
\noindent {\it Proof.} The existence follows from the explicit formula given
below in terms of Riemann Theta functions. Uniqueness results from the
Riemann--Roch theorem applied to the ratio of two such functions. \square

We now give a fundamental formula expressing the Baker--Akhiezer functions
in terms of Riemann theta functions.
According to the Riemann-Roch theorem for any divisor
$D=\gamma_1+\cdots+\gamma_{g+l-1}$ in general position there exists a unique
meromorphic function $h_{\alpha}(P)$ such that the divisor of
its poles coincides with $D$ and such that
\beq
h_{\alpha}(P_j)=\delta_{\alpha j}. \label{2.26}
\eeq
Using the results recalled in Appendix B, this function may be written
as follows:
\beq
h_{\alpha}(P)={f_{\alpha}(P)\over f_{\alpha}(P_{\alpha})}; \quad
f_{\alpha}(P)=\theta(A(P)+Z_{\alpha})
{\prod_{j\neq \alpha}\theta(A(P)+R_j)\over \prod_{i=1}^l\theta
(A(P)+S_i)}, \label{2.260}
\eeq
where
\beq
R_j=-{\cal K}-A(P_j)-\sum_{s=1}^{g-1} A(\gamma_s), \ \ j=1,\ldots,l
\eeq
\beq
S_i=-{\cal K}-A(\gamma_{g-1+i})-\sum_{s=1}^{g-1} A(\gamma_s),
\eeq
and
\beq
Z_{\alpha}=Z_0-A(P_{\alpha}), \quad
Z_0=-{\cal K}-\sum_{i=1}^{g+l-1} A(\gamma_i)+\sum_{j=1}^l A(P_j).
\label{2.263}
\eeq

Let $d\Omega^{(i)}$ be the unique meromorphic
differential holomorphic on $\G$ outside the punctures $P_j,\
j=1,\ldots,l$, which has the form
\beq
d\Omega^{(i)}=d(w_j^{-i}+O(w_j)) \label{2.23}
\eeq
near the punctures and is normalized by the conditions (see Appendix B
for the definition of the cycles $a^0_k,~ b^0_k$)
\beq
\oint_{a_k^0}d\Omega^{(i)}=0. \label{2.24}
\eeq
It defines a vector $U^{(i)}$
with coordinates
\beq U^{(i)}_k={1\over 2\pi i} \oint_{b_k^0} d\Omega^{(i)} \label{2.25}
\eeq
\begin{th}
The components of the Baker-Akhiezer function
$\psi(x,t,P)$ are equal to
\beq
\psi_{\alpha}(x,t,P)=h_{\alpha}(P) {\theta
(A(P)+U^{(1)}x+U^{(2)}t+Z_{\alpha}) \theta (Z_0) \over \theta (A(P)+Z_{\alpha})
\theta (U^{(1)}x+U^{(2)}t+Z_0) } e^{(x\Omega^{1}(P)+t\Omega^{(2)}(P))}
\label{2.29}
\eeq
\beq \Omega^{(i)}(P)=\int_{q_0}^P d\Omega^{(i)} \label{2.30}
\eeq
\end{th}

\noindent{\it Proof.} It is enough to check that the
function defined by the formula (\ref{2.29}) is well-defined
( i.e. it does not
depend on the path of integration between $q_0$ and $P$) and has the desired
analytical properties. (The ratio of the product of theta-functions at
$P=P_{\alpha}$ equals 1 due to (\ref{2.263}).)
\square

 From the exact theta-function formula~(\ref{2.29}) it follows
that $\psi$ may be represented in the form
\beq
\psi(x,t,P)=r(x,t)\hat \psi(x,t,P), \label{2.100}
\eeq
where $\hat \psi(x,t,P)$ is an entire function of the variables $x,t$ and
$r(x,t)$ is meromorphic function.

We now give the definition of the dual Baker--Akhiezer function.
For any set of $g+l-1$ points in general position there exists a unique
meromorphic differential $d\Omega$ that has poles of the form
\beq
d\Omega={dw_j\over w_j^2}+{\lambda_j dw_j\over w_j}+O(1)dw_j
\label{2.3a}
\eeq
at the punctures $P_j$, and equals zero at the points $\gamma_s$
\beq
d\Omega(\gamma_s)=0. \label{2.3b}
\eeq
Besides $\gamma_s$ this differential has $g+l-1$ other zeros that we denote
$\gamma_s^+$.

The dual Baker-Akhiezer function is the unique function $\psi^+(x,t,P)$ with
coordinates $\psi^{+,\alpha}(x,t,P)$ such that

$1^0.$ The function $\psi^{+,\alpha}$ as a function of the variable
$P\in \Gamma$ is meromorphic outside the punctures and has at most
simple poles at the points  $\gamma_s^+$ (if all of them are distinct);

$2^0.$ In the neighbourhood of the puncture $P_j$ it  has the form
\beq
\psi^{+,\alpha}(x,t,P)=e^{-w_j^{-1}x-w_j^{-2}t}\left(\delta_{\alpha j}+
\sum_{s=1}^{\infty}\xi_s^{+,\alpha j}(x,t)w_j^{s}\right).\label{2.1a}
\eeq

Let $h^+_{\alpha}(P)$ be the function that has poles at the points of the dual
divisor $\gamma_1^+,\ldots,\gamma_{g+l-1}^+$ and is normalized by (\ref{2.26})
(i.e. $h_{\alpha}^+(P_j)=\delta_{\alpha j}$). It can be written in the form
(\ref{2.260}) in which $\gamma_s$ are replaced by $\gamma_s^+$.
It follows from the definition of the dual divisors  that
\beq
\sum_{s=1}^{g+l-1} A(\gamma_s)+\sum_{s=1}^{g+l-1}
A(\gamma_s^+)=K_0+2\sum_{j=1}^l A(P_j), \label{2.265}
\eeq
where $K_0$ is the canonical class (i.e. the equivalence class of
of the divisor of zeros of a holomorphic differential).

\begin{th}
The components of the dual Baker-Akhiezer function
$\psi^+(x,t,P)$ are equal to
\beq
\psi_{\alpha}^+(x,t,P)=h_{\alpha}^+(P)
{\theta (A(P)-U^{(1)}x-U^{(2)}t+Z_{\alpha}^+) \theta (Z_0^+)
\over \theta (A(P)+Z_{\alpha}^+) \theta (U^{(1)}x+U^{(2)}t-Z_0^+) }
e^{-(x\Omega^{1}(P)+t\Omega^{(2)}(P))}, \label{2.290}
\eeq
where
\beq
Z_0^+=Z_0-2{\cal K}-K_0 , \ \ \
Z_{\alpha}^+=Z_0^+-A(P_{\alpha}). \label{2.291}
\eeq
\end{th}

The above results are valid for any curve $\G$. We now consider a more
specific setting which will correspond to the elliptic model.
\begin{th}
\label{specific}
Let $\Gamma$ be a smooth algebraic curve defined by an equation
of the form
\beq
R(k,z)=k^N+\sum_{i=0}^N r_i(z)k^i=0, \label{2.7}
\eeq
where $r_i(z)$ are elliptic functions, holomorphic outside the point $z=0$,
such that the covering $P \to z$ has no branching points over $z=0$ (i.e.
the function $k(P)$ has $N$ simple poles $P_1,\ldots,P_N$ on $\Gamma$
which are
preimages of $z=0$). Let us assume also that the residues $\nu_j$ of $k(P)$
at the poles defined by the expansion of $R(z,k)$ near $z=0$
\beq
R(k,z)=\prod_{i=1}^N(k+\nu_i z^{-1}+O(z^0)), \label{2.8}
\eeq
satisfy (\ref{1.30}), $\nu_j=1, \ j>l$. Then there exists a function
$\varphi_i(P)$ on $\G$ such that the Baker--Akhiezer function $\psi$
corresponding to the curve $\G$ and the local parameters
$w_j=(k_j(z)+\zeta(z))^{-1}$ at the puncture $P_j$ obeys:
\beq
\psi(x+2\omega_i,t,P)=\varphi_i(P)\psi(x,t,P). \label{2.11}
\eeq
\end{th}

{\it Proof.} Consider the functions
\beq
\varphi_i(P)=\exp(2(k(P)+\zeta(z))\omega_i-2\eta_iz),\ \ i=1,2, \label{2.9}
\eeq
where $2\omega_i$ are periods of the elliptic base curve and
$\eta_i=\zeta(\omega_i)$. From the monodromy properties of
$\zeta$-function
and relation $2(\eta_1\omega_2-\eta_2\omega_1)=\pi i$ it follows
that $\varphi(P)$
is a well-defined function on the curve $\Gamma$. It is holomorphic
outside the
points $P_1,\ldots,P_l$. In the neighbourhood of $P_j$ it has the form
\beq
\varphi_i(P)=(1+O(z))\exp(2(k_j(z)+\zeta(z))\omega_i). \label{2.10}
\eeq
Let $\psi(x,t,P)$ be the Baker-Akhiezer vector function corresponding to
$\Gamma, P_j, w_j(P)$ and to any divisor $D$ of the degree $g+l-1$. Then
equation~(\ref{2.11}) follows from the fact that the right and left
hand sides have the same analytical properties. \square

\begin{cor}
The vector Baker--Akhiezer function $\psi(x,t,P)$ with components
$\psi_\alpha, \; \alpha =1,\cdots, l$ can be written in the form:
\beq
\psi(x,t,P)=\sum_{i=1}^m s_i(t,P)\Phi(x-x_i(t),z) e^{kx+k^2t} , \quad
P=(k,z), \label{2.12}
\eeq
\end{cor}

\noindent{\it Proof.}
Let $x_i(t), i=1,\ldots,m,$ be the set of poles of the function $\psi(x,t,P)$
(as a function of the variable $x$) in the fundamental domain of
the lattice with periods $2\omega, 2\omega'$.
It follows from (\ref{2.100}) that
they do not depend on $P$. Let us assume that these poles are simple.
Then their exist vectors $s_i(t,P)$ such that the function
$$
{\cal F}(x,t,P)=\psi(x,t,P)-\sum_{i=1}^m s_i(t,P)\Phi(x-x_i(t),z)
e^{kx+k^2t}
$$
is {\it holomorphic} in $x$ in this fundamental domain. This function has the
same
monodromy properties (\ref{2.11}) as the function $\psi$.
Any non-trivial function satisfying (\ref{2.11}) has at least one pole in
the fundamental domain. Hence, ${\cal F}=0$. \square

Let us remark that for the above specific curve $\G$, the
Baker--Akhiezer function is exactly of the form postulated in
equation~(\ref{1.4}).
The same arguments show that the dual Baker-Akhiezer function
has the form~(\ref{1.4a}).

\subsection{The potential.}

The following theorem is a particular case of a general statement
\cite{kr3}.
\begin{th}
Let $\psi(x,t,P)$ be a vector-function with
above-defined co-ordinates \\ $\psi_{\alpha}(x,t,P)$.
Then it satisfies the equation
\beq (\p_t-\p_x^2+u(x,t))\psi(x,t,P)=0,\label{2.2}
\eeq
where the entries of the matrix-function $u$ are equal to:
\beq
u^{\alpha i}(x,t)=2\p_x\xi_1^{\alpha i}(x,t).\label{2.3}
\eeq
The potentials $u(x,t)$ corresponding to some Baker-Akhiezer
vector-function are
called algebro-geometrical or finite-gap potentials.
\end{th}

\noindent{\it Proof.} This follows directly from the fact that
$(\p_t-\p_x^2)\psi_\alpha$ has the same analytic properties as $\psi$ but
for the normalizations at the $P_i$'s, so can be expanded on
the $\psi_\beta$ with coefficients $-u^{\alpha\beta}$. \square

\begin{th}
The dual Baker-Akhiezer function satisfies  the equation:
\beq
\psi^+(x,t,P)(\p_t-\p_x^2+u(x,t))=0,\label{2.2a}
\eeq
where $u(x,t)$ is the same as in (\ref{2.2}).
\end{th}

\noindent{\it Proof.}
To show that the potentials are the same, we consider the form
$\psi_\alpha \psi^{+\beta}d\Omega$ where
$d\Omega$ is defined by equation~(\ref{2.3a}) and the conditions~(\ref{2.3b}).
This is a meromorphic
1--form on $\G$ with poles only at the $P_j$'s. Around $P_j$ we have:
$$\psi_\alpha \psi^{+\beta}d\Omega=\left[
{\delta_{\alpha j}\delta_{\beta j}\over w_j^2}+(\delta_{\alpha
j}\delta_{\beta j}\lambda_j+\delta_{\alpha
j}\xi_1^{+\beta_j}+\delta_{\beta j}\xi_1^{\alpha j}){1\over
w_j}+O(1)\right] dw_j$$
Since the sum of residues must vanish we get:
$$\xi_1^{+\beta\alpha}+\xi_1^{\alpha\beta}= -\lambda_\alpha
\delta_{\alpha\beta}$$
This implies the result since $u^{\alpha\beta}=2\p_x\xi_1^{\alpha\beta}$,
$u^{+\alpha\beta}=-2\p_x\xi_1^{+\beta\alpha}$ and
$\lambda_\alpha$ is independent of $x$. \square

In general position the algebro-geometrical potentials are quasi-periodic
functions of all arguments. In \cite{kr2} the necessary conditions on the
algebraic geometrical data \\
$\{\Gamma,P_1,\ldots,P_l,w_1(P),\ldots,w_l(P)\}$ were found in order
that the corresponding potentials be elliptic functions
of the variable $x$. Here we have

\begin{prop}
Let $\G$ be a specific curve as in Theorem~(\ref{specific}).
Then the
algebro-geometrical potential $u(x,t)$ corresponding to
the curve $\Gamma$, to the points $P_1,\ldots,P_l$, and to the local
coordinates
$w_j(z)=(k_j(z)+\zeta(z))^{-1}$  is an elliptic function. In general position
this potential has the form
\beq
u(x,t)=\sum_{i=1}^N a_i(t)b_i^+(t)\wp(x-x_i(t)) \label{2.81}
\eeq
\end{prop}

\noindent{\it Proof.} The potential is elliptic because of
equation~(\ref{2.11}).
Since the Baker--Akhiezer function has the form~(\ref{2.12})
$u$ has only double poles. Hence it is of
the form: $u=\sum \rho_i(t) \wp(x-x_i(t))$. We show that the matrices
$\rho_i(t)$ are of rank one.
It follows from (\ref{2.29}) that the poles $x=x_i(t)$ of the Baker-Akhiezer
function
correspond to the solutions of the equation:
\beq
\theta(U^{(1)}x+U^{(2)}t+Z)=0. \label{spain3}
\eeq
 From (\ref{2.263}) it follows that for such a pair $( x_i(t),\ t)$
the first factor in the numerator of formula (\ref{2.29}) vanishes
at $P_\alpha$. At a point $P_\beta$, $\beta \neq \alpha$, it is the
function $h_\alpha(P)$ which vanishes.
Therefore, the residue $\psi^0_{\alpha,i}(t,P)$ of
$\psi_{\alpha}(x,t,P)$ at $x=x_i(t)$, as a function of the variable $P$,
has the following analytical properties:

$1^0.$ It is a meromorphic function outside the punctures $P_j$
and has the same set of poles as $\psi$;

$2^0.$ In a neighbourhood of the puncture $P_j$ it has the form
\beq
\psi^0_{\alpha,i}(t,P)=\exp(w_j^{-1}x_i(t)+w_j^{-2}t)O(w_j). \label{spain4}
\eeq
Hence, it has the same analytical properties as
the Baker--Akhiezer function but with one difference. Namely, the
regular factor of its expansion at {\em all} the punctures vanishes.
For generic $x,t$ there are no such function. For special pairs
$(x=x_i(t),t)$ such a function $\psi_{i0}(t,P)$
exists and is {\it unique} up to a constant ( in $P$ ) factor
(it is unique in a generic case when $x_i(t)$ is a simple root of
(\ref{spain3})).
Therefore, the components of the Baker-Akhiezer function can be
represented in the form
\beq
\psi_{\alpha}(x,t,P)={\phi_{\alpha}(t)\;\psi_{i0}(t,P)\over x-x_i(t)}+
O((x-x_i(t))^0).
\label{spain5}
\eeq
The last equality implies that the residue $\tilde \rho_i(t)$ of the matrix
$\xi_1(x,t)$ with entries $\xi_1^{\alpha j}(x,t)$
\beq
\xi_1(x,t)={\tilde \rho_i(t)\over x-x_i(t)} +O((x-x_i(t))^0)
\eeq
is of rank 1.
It follows from (\ref{2.3}) that
$\rho_i(t)=-2\tilde \rho_i(t)$. Hence this matrix is of rank one too,
and therefore has the form (\ref{2.81}). \square

Let us examine now the effect of replacing the divisor by an
equivalent one.
Let $D=\gamma_1+\cdots+\gamma_{g+l-1}$ and
$D^{(1)}=\gamma_1^{(1)}+\cdots+\gamma_{g+l-1}^{(1)}$ be two equivalent divisors
(i.e. there exists a meromorphic function $h(P)$ on $\Gamma$ such
that $D$ is a divisor of its poles and $D^{(1)}$ is a divisor of its zeros).
Then for the corresponding Baker-Akhiezer vector-functions
$\psi(x,t,P)$ and $\psi^{(1)}(x,t,P)$ the equality
\beq
H\psi(x,t,P)=\psi^{(1)}(x,t,P)h(P), \label{2.4}
\eeq
is valid. Here $H$ is a diagonal matrix
\beq
H^{\alpha j}=h(P_j)\delta^{\alpha j}. \label{2.5}
\eeq
The proof of (\ref{2.4}) follows from the uniqueness of the Baker-Akhiezer
functions, because the left and the right hand sides of (\ref{2.4}) have the
same analytical properties.

\begin{cor}
The algebraic-geometrical potentials $u(x,t)$ and $u^{(1)}(x,t)$ corresponding
to $\Gamma,P_j,w_j(P)$ and to equivalent effective divisors $D$ and $D^{(1)}$,
respectively, are gauge equivalent
\beq
u^{(1)}(x,t)=Hu(x,t)H^{-1}, H^{\alpha j}=h_j\delta^{\alpha j}. \label{2.6}
\eeq
\end{cor}

\begin{cor}
A curve $\Gamma$ in a general position  satisfies the
conditions of theorem 6.2 if and only if it is the spectral
curve~(\ref{1.25}) of a Lax matrix $L$ defined by the formula
(\ref{1.17}) where
$x_i, p_i$ are arbitrary constants and $f_{ij}$ are defined by vectors
$a_i, \ b_i^+$, satisfying (\ref{7}), with the help of (\ref{8}).
The corresponding
vectors are defined uniquely up to the transformation (\ref{1.52})
\end{cor}

Notice that the Baker--Akhiezer function
$\Psi_\alpha(x,t,P)/\Psi_1(0,0,P)$ appearing in equation~(\ref{1.39})
is related to the normalized Baker--Akhiezer function
$\psi_\alpha(x,t,P)$ appearing in equation~(\ref{2.1}) by:
$${\Psi_\alpha(x,t,P)\over \Psi_1(0,0,P)}=\sum_\beta \chi_0^{\alpha\beta}
\psi_\beta(x,t,P)$$
This induces a similarity transformation on the potential, and we have:

\begin{cor}
If $a_i(t),b_i(t),x_i(t)$ are a solution of the equation of motion of the
hamiltonian system (\ref{1}) then
\beq
\sum_{i=1}^N a_i (t) b_i^+(t)\wp(x-x_i(t)) = \chi_0 u(x,t)\chi_0^{-1},
\label{2.16}
\eeq
where $u(x,t)$ is the algebro--geometrical potential corresponding to the
data that are described in theorem 6.2 and to the normalized Baker--Akhiezer
functions.
\end{cor}
\begin{cor}
The correspondence
\beq
a_i(t),\ b_i^+(t), x_i(t) \longmapsto \{\G,\ [D]\},
\label{2.16b}
\eeq
where $[D]$ is a equivalence class of the divisor $D$ (i.e. the corresponding
point of the Jacobian) is an isomorphism up to the equivalence (\ref{1.52}).
\end{cor}
The curve $\G$ is a time independent. At the same time $[D]$ depends on the
choice of the normalisation point $t_0=0$. From the exact formulae for the
solutions (see below) it follows that the dependence of
$D(t_0)$ is linear onto the Jacobian.

\subsection{Reconstruction formulae.}

\begin{th} Let $\G$ be a curve that is defined by the equation of the form
(\ref{1.25}) and $D=\gamma_1,\ldots,\gamma_{g+l-1}$ be a set of points in
general position. Then the formulas
\beq
\theta (U^{(1)}x_i(t)+U^{(2)}t+Z_0)=0,  \label{2.31}
\eeq
\beq
a_{i, \alpha}(t)=Q_i^{-1}(t)h_{\alpha}(q_0)
{\theta (U^{(1)}x_i(t)+U^{(2)}t+Z_{\alpha}) \over
\theta (Z_{\alpha})} . \label{2.32}
\eeq
\beq
b_i^{\alpha}(t)=Q_i^{-1}(t)h_{\alpha}^+(q_0)
{\theta (U^{(1)}x_i(t)+U^{(2)}t-Z_{\alpha}^+) \over
\theta (Z_{\alpha}^+)} , \label{2.33}
\eeq
where
\beq
Q_i^2(t)={1\over 2} \sum_{\alpha=1}^lh_{\alpha}^+(q_0) h_{\alpha}(q_0)
{\theta (U^{(1)}x_i(t)+U^{(2)}t-Z_{\alpha})
 \theta (U^{(1)}x_i(t)+U^{(2)}t-Z_{\alpha}^+) \over \theta (Z_{\alpha})
\theta (Z_{\alpha}^+)}    \label{2.35}
\eeq
define the solutions of the system (\ref{4}, \ref{1.6}, \ref{1.7}). Any
solution
of the system (\ref{1}) may be obtained from the solutions
(\ref{2.31}-\ref{2.33}) with the help of symmetries (\ref{1.52}).
\end{th}

{\bf Remark.} (\ref{2.31}) may be reformulated as follows: the vector $U^{(1)}$
defines an imbedding of the elliptic curve into the Jacobian $J(\G)$. Therefore
the restriction of the theta function onto the corresponding elliptic curve is
a product of the elliptic $\sigma$-Weiestrass functions, i.e.
of the vector
\beq
\theta (U^{(1)}x+U^{(2)}t+Z_0)=const \prod_{i=1}^N \sigma (x-x_i(t)).
\label{2.310}
\eeq
It should be mentioned that exactly the same formula was obtained in
\cite{kr2} for the solutions of the elliptic Calogero-Moser system. The
difference
for various spins $l$ is encoded in the set of corresponding curves.

 From (\ref{1.4},\ref{1.13}) it follows that the vector $a_i(t)$ is
proportional
to the singular part of the expansion of $\psi(x,t,P)$ near $x=x_i(t)$ and up
to
the scalar factor this singular part does not depend on $P$. Therefore, using
the formula (\ref{2.29}) for $P=q_0$ we obtain (\ref{2.32}).\square

\section{The rational and trigonometric cases.}

\subsection{Baker--Akhiezer functions}
Let $\Gamma$ be the smooth algebraic curve of genus $g$ defined by the
equation~(\ref{gammatrig}) or (\ref{gammarat}). The function $k(P)$ has $l$
simple poles
above $z=0$, denoted by $P_1,\cdots,P_l$. In the vicinity of $P_i$ one
has $k=k_i(z)\equiv -\nu_i/z-h_i(z)$ with $h_i$ regular. We take
$1/k_i(z)$ as local parameter around $P_i$.
We shall consider the space of Baker--Akhiezer functions with
$(g+l-1)$ poles $\gamma_k$ and $N$ poles at the points $Q_j$ above
$z=-\infty$,
which are of the form $\exp\,(kx+k^2t)$ times a
meromorphic function. The space of such functions
is of dimension $N+l$, and imposing the $N$ conditions~(\ref{babeltrig}) or
(\ref{babel})
one ends up with a space of dimension $l$. One can define a
basis $\psi_\alpha$ of this space
by choosing the normalization in the neighbourhood of
the points $P_j$ above $z=0$ with local parameters $(z,k_j)$
as:
\beq
\psi_\alpha(P,x,t)=\left(\delta_{\alpha j}+\sum_{s=1}^\infty
{\xi_s^{\alpha j}(x,t) \over k_j^s}\right) e^{k_j x+k_j^2
t}\label{baker}
\eeq

We have now at our disposal $g+l-1$ parameters $\gamma_k$, and the $N$
coefficients $\mu_j$ which add up to $m+l$, i.e. half of
the dimension of the phase space, just as in the elliptic case.

One can construct directly the Baker functions $\psi_\alpha$ from the
above analyticity properties. One first chooses a convenient basis to
expand them.

There exists a unique function $g_j(P)$ with $g+l-1$ poles
$\gamma_k$, one pole at $Q_j$
with residue 1 (i.e. of the form $1/(k-\chi_j)$)
and $l$ zeroes at the $P_\alpha$'s. Then we can write:
\begin{equation}
\psi_\alpha(P,x,t)=\left( h_\alpha(P)+\sum_{j=1}^N g_j(P) r_{j \alpha }(x,t)
\right)e^{kx+k^2t}\label{psialp}
\end{equation}
where $h_\alpha(P)$ is the function defined in eq.(\ref{2.260}),
so that conditions~(\ref{baker}) are fulfilled.

It remains only to express the $N$ conditions~(\ref{babeltrig} or \ref{babel}).
This yields the

\begin{th}
The components of the Baker--Akhiezer function $\psi(x,t,P)$ are given by:
\begin{equation}
\psi_\alpha(x,t,P)=\, {{\rm Det}
\pmatrix{h_\alpha(P)&g_j(P)\cr h_\alpha(T_i)&\Theta_{ij}(x,t)}\over
{\rm Det}\pmatrix{\Theta_{ij}(x,t)}}\, e^{kx+k^2t}\label{belleformule}
\end{equation}
where $\Theta$ is the matrix with elements
in the trigonometric case:
\beq
\Theta_{ii}=-\sigma_i e^{-2x- 4(\chi_i +1)t} + g_i(T_i), \quad
\Theta_{ij}= g_j(T_i)\quad i\neq j . \label{tetatrig}
\eeq
In the rational case one has to replace $h_\alpha(T_i)$ by
$h_\alpha(Q_i)$ in~(\ref{belleformule}) and to define:
\beq
\Theta_{ii}=x+ 2\chi_i t - \sigma_i+g^{(1)}_i, \quad
\Theta_{ij}= g_j(Q_i)\quad i\neq j \label{teta}
\eeq
where $g_j(P)=1/(k-\chi_j)+ g^{(1)}_j+O(k-\chi_j)$.
\end{th}

\noindent {\it Proof.}
We first express the conditions on $\psi$ arising from the the
conditions~(\ref{babeltrig}) or~(\ref{babel}) on $\psi/f$ where $f$ is
the meromorphic function introduced in equation~(\ref{funf}).
As a matter of fact near the
point $Q_j$ in the trigonometric case we have:
$$\psi_\alpha(x,t,P)={R^{(-1)}_{j\alpha }(x,t)\over
k-\chi_j}+O((k-\chi_j)^0)$$
while around $T_j$ we have:
$$\psi_\alpha(x,t,P)=R^{(0)}_{j\alpha }(x,t)+O(k-\chi_j)$$
The relations~(\ref{babeltrig}) take the form:
\beq
R^{(0)}_{j\alpha }(x,t)=
\sigma_j  R^{(-1)}_{j\alpha}(x,t)
\quad {\rm with~}
\sigma_j=\mu_j {f(T_j) \over f^{(0)}_j } \label{newbabeltrig}
\eeq
where around $Q_j$ the function $f$ appearing in equation~(\ref{funf})
has the corresponding expansion:
$$f(P)={f^{(0)}_j\over k-\chi_j}+f^{(1)}_j+O(k-\chi_j)$$ and $\sigma_j$ is
independent of $x$ and $t$.
In the rational case near $Q_j$ we have:
$$\psi_\alpha(P,x,t)={R^{(-1)}_{j\alpha }(x,t)\over
k-\chi_j}+R^{(0)}_{j\alpha }(x,t)+O(k-\chi_j)$$
and the conditions~(\ref{babel}) on $\psi$ are equivalent to:
\beq
R^{(0)}_j(x,t)=\sigma_j R^{(-1)}_j(x,t)\quad {\rm with~}
\sigma_j=\mu_j+{f^{(1)}_j\over f^{(0)}_j} \label{newbabel}
\eeq

Using the expression~(\ref{psialp}) of $\psi$ these conditions take
the form (in the rational case $T_j$ is replaced by $Q_j$ below):
\begin{equation}
\sum_k \Theta_{jk}(x,t)r_{k\alpha}=-h_{\alpha}(T_j)
\end{equation}
Solving this linear system with Cramer's rule yields the result.
\square

\begin{prop}
The Baker--Akhiezer function given in~(\ref{belleformule}) can be
written in the form:
\beq
\psi=\sum_i^Ns_i(t,k,z) \Phi ( x-x_i(t), z)
e^{kx+k^2t}\label{bfd}
\eeq
\end{prop}

\noindent{\it Proof.}
Let us give the proof in the trigonometric case. In the rational case, the
proof is similar and even simpler. From eq.(\ref{belleformule}) we see that
one can write
\begin{eqnarray}
\psi_\alpha (x,t,P)& =& \left( h_\alpha(P) - \sum_{i=1}^N
 {2 e^{-2x_i} s_{i,\alpha}(t,P) \over {e^{-2x} -e^{-2x_i}}}
\right)e^{kx + k^2 t}
\nonumber \\
&=&  \left( h_\alpha(P) + \sum_{i=1}^N s_{i,\alpha}(t,P)
+  \sum_{i=1}^N s_{i,\alpha}(t,P) \coth(x-x_i)\right) e^{kx + k^2 t}
\label{psidecompose}
\end{eqnarray}
We have to show that
\begin{eqnarray}
h_\alpha (P) = -( 1 + \coth z) \sum_{i=1}^N s_{i,\alpha}(t,P)
\nonumber
\end{eqnarray}
But the function $h_\alpha(P) /(1 + \coth z)$ vanishes at the points
$P_i$ above $z=0$, and has poles at the points $Q_j$ above $z=-\infty$.
Hence, we can write
\begin{eqnarray}
{h_\alpha(P) \over {1+\coth z}} = \sum_{j=1}^N
{h_\alpha(Q_j) \over \alpha_j } g_j(P)
\nonumber
\end{eqnarray}
where the constants $\alpha_j$ are defined by $1+\coth z= \alpha_j\,
 (k-\chi_j) +
O(k-\chi_j)^2$ around $Q_j$. Using this formula at $P=T_i$ we get
in particular
\begin{eqnarray}
h_\alpha(T_i) = 2 \sum_{j=1}^N {h_\alpha(Q_j) \over \alpha_j }
g_j(T_i)
\label{haq}
\end{eqnarray}
On the other hand, using eq.(\ref{psidecompose}), we find
\begin{eqnarray}
2\sum_{i=1}^N s_{i,\alpha}(t,P) = \psi_\alpha(x,t,P)e^{-kx-k^2t}
\vert_{x=+\infty} ~-~  \psi_\alpha(x,t,P)e^{-kx-k^2t}\vert_{x=-\infty}
= {{\rm Det}\, \pmatrix{0  & g_j(P) \cr h_\alpha(T_i) &
g_j(T_i) } \over {\rm Det}\, \pmatrix{ g_j(T_i)}}
\nonumber
\end{eqnarray}
Expanding the determinant in the numerator along the first line, and using
eq.(\ref{haq}) to evaluate $h_\alpha(T_j)$, we get
\begin{eqnarray}
\sum_{i=1}^N s_{i,\alpha}(P,t) =-
\sum_{j=1}^N {h_\alpha(Q_j) \over \alpha_j} g_j(P)
\nonumber
\end{eqnarray}
which is what we had to prove.
\square

One can also give an explicit formula for $s_{i,\alpha}(t,P)$.
Since
${\rm Det}\, \Theta(x_i,t)=0$ one can write a linear dependency
relation:
$$\Theta_{k1}=\sum_{j=2}^{N}\lambda_j^{(i)}(t) \Theta_{kj},\quad\forall k$$
and we see that the residue $s_{i, \alpha}(t,P)$ in eq.~(\ref{psidecompose})
is given by ($\lambda_1^{(i)}=-1$):
\beq
s_{i ,\alpha}(t,P)=\left\{ {\sum_{k=1}^N
\lambda_k^{(i)}(t)g_k(P) \over \prod_{j=1}^N
(2\sigma_je^{-4(\chi_j+1)t-x_i-x_j})
\prod_{j\neq i}\sinh\,(x_i-x_j)}
\right\}    \,{\rm
Det}\, \Theta_\alpha^{(i)}(t) \label{sitp}
\eeq
Here $\Theta_\alpha^{(i)}$ is obtained from $\Theta(x,t)$ by taking
$x=x_i$ and replacing the first column by $h_\alpha(T_j)$.
This equation has  to be compared with eq.(\ref{1.13}).
In the
rational case we find a similar and simpler formula, including the
same factor $\Theta_\alpha^{(i)}(t)$.

 As in the elliptic case we need the dual Baker--Akhiezer function
$\psi^+$ and for this we introduce the differential $d\Omega$ with
poles of order 2 at the punctures $P_j$'s such that $d\Omega=
dw_j/w_j^2+O(1/w_j)dw_j$ and vanishing on the $g+l-1$ points
$\gamma_k$. Let $\gamma_k^+$ the $g+l-1$ other zeroes of $d\Omega$.

Let $h^{+,\alpha}(P)$ be the unique function with poles at the
$\gamma_k^+$'s  and such that $ h^{+,\alpha}(P_j)=\delta_{\alpha j}$.
In the trigonometric case we introduce the function
 $g^+_j(P)$  with $g+l-1$ poles
$\gamma^+_k$, one pole at $T_j$ with residue 1
(i.e. of the form $1/(k-\chi_j-2)$),
and $l$ zeroes at the $P_j$'s. Then we define the dual
Baker--Akhiezer function:
\begin{equation}
\psi^{+,\alpha}(P,x,t)=\left( h^{+,\alpha}(P)+\sum_{j=1}^N g^+_j(P)
r_j^{+,\alpha }(x,t)
\right)e^{-kx-k^2t}.\label{psialp+}
\end{equation}
and such that relations of the type (\ref{newbabel}) are satisfied
with some coefficients $\sigma_j^+$.
We choose $\sigma_j^+$ as
\begin{eqnarray}
\sigma_j^+ = -\sigma_j  {d\Omega (T_j) \over d\Omega (Q_j)}
\nonumber
\end{eqnarray}
where the form $d\Omega$ is expressed on $dk$ at $Q_j$ and $T_j$. With this
choice the sum of the residues of $\psi^{+,\alpha}\psi_\beta
d\Omega$ at $Q_j$ and $T_j$ vanishes. This condition ensures that the
potential reconstructed from $\psi^+$ is the same as the one
reconstructed from $\psi$.

Notice that the roles of $Q_j$ and $T_j$ are interchanged in the definitions
of $\psi$ and $\psi^+$.

A similar analysis holds in the rational case.

\begin{th}
The components of the Baker--Akhiezer function $\psi^+(x,t,P)$ are given by:
\begin{equation}
\psi^{+,\alpha}(P,x,t)=\, {{\rm Det}
\pmatrix{h^{+,\alpha}(P)&g^+_j(P)\cr h^{+,\alpha}(Q_i)&\Theta^+_{ij}(x,t)}\over
{\rm Det}\pmatrix{\Theta^+_{ij}(x,t)}}\, e^{-kx-k^2t}\label{belleformule+}
\end{equation}
where $\Theta^+$ is the matrix with elements:
\beq
\Theta^+_{ii}=-\sigma_i^+ e^{2x+4(\chi_i+1)t}+g_i^+(Q_i),\quad
\Theta^+_{ij}=g_j^+(Q_i).
\eeq
In the rational case one has to define:
\beq
\Theta^+_{ii}=- x- 2\chi_i t -\sigma_i^+ + g_i^{(1)+}, \quad
\Theta^+_{ij}= g^+_j(Q_i). \label{teta+}
\eeq
\end{th}

\subsection{The potential.}

\begin{th}
The vector Baker--Akhiezer function $\psi(x,t,P)$ is a solution of the
equation $(\p_t-\p^2_x+u(x,t))\psi=0$ where the potential $u$ is given
by: $u(x,t)=\sum \rho_i(t)V(x-x_i(t))$ and $\rho_i(t)$ is an $l\times l$
matrix of rank 1.
\end{th}

\noindent{\it Proof.} The usual argument from the unicity of the
Baker--Akhiezer function shows that $(\p_t-\p^2_x)\psi$ is of the form
$-u(x,t)\psi$ with $u=2\p_x\xi_1^{\alpha j}(x,t)$. From
equation~(\ref{bfd}) it is clear that $u(x,t)$ is of the form
$\sum \rho_i(t)V(x-x_i(t))$. To compute $\rho_i$ let us expand around $P_\beta$
$$g_i(P)={g_i^\beta\over k_\beta}+O({1\over k_\beta^2})\quad{\rm and}\quad
h_\alpha(P)=\delta_{\alpha \beta}+{h_\alpha^\beta\over k_\beta}+O({1\over
k_\beta^2})$$
and $s_{i,\alpha}(t,P)$ given in eq.~(\ref{sitp}).
We find $\rho_{i,\alpha}^\beta=a_{i,\alpha}b_i^\beta$ where:
$$a_{i,\alpha}={ 1\over Q_i(t)}{\rm Det}\,\Theta_\alpha^{(i)}(t) \quad
b_i^\beta=-2 Q_i(t)
\left\{ {\sum_{k=1}^N
\lambda_k^{(i)}(t)g_k^\beta \over
\prod_{j=1}^N (2\sigma_je^{-4(\chi_j+1)t-x_i-x_j})
\prod_{j\neq i}\sinh\,(x_i-x_j)}
\right\}
$$\square

Alternatively one could use the dual Baker function $\psi^+$. It
satisfies a Schr\"odinger equation with the same potential $u$ than
$\psi$. This is because the sum of the residues of the form $\psi^{+\alpha}
\psi_\beta \Omega$ at the points $Q_j$ and $T_j$ vanishes, so that the
same argument
as in the elliptic case applies. This shows that $\Theta$ and
$\Theta^+$ have the same eigenvalues $-x_i(t)$ and gives alternative
formulae for $a_i^\alpha$ and $b_i^\beta$, in particular:
$$b_i^\alpha ={1\over Q^+_i(t)}{\rm Det}\,\Theta^{+,\alpha (i)}(t)$$
The normalizations $Q_i(t)$ and $Q_i^+(t)$ are as usual determined by
the conditions $f_{ii}=2$ and $\sum_\alpha b_i^\alpha=1$.

This implies that $x_i(t), a_i(t), b_i(t)$ are the solutions of the
 trigonometric or rational  model. Note that the curve is
necessarily of the form of the spectral curve of the Calogero model.

\subsection{Reconstruction formulae.}

To construct the functions $g_j$ one can take advantage of the fact
that we know on $\G$ the function $1/(k-\chi_j)$ which vanishes at the
$l$ punctures $P_\alpha$ and has a pole with residue 1 at $Q_j$. It
has $l-1$ other poles at some well--defined points
$\delta^{(j)}_k$. The function $g_j(P)(k-\chi_j)$ has $g+l-1$ poles
$\gamma_k$ and $l-1$ zeroes $\delta^{(j)}_k$. By Riemann--Roch theorem
this function $F_j(P)$
is uniquely determined by these data and the normalization condition
$F_j(Q_j)=1$. One can give an expression in terms of theta
functions as in eq.(\ref{2.260}). Then
$$g_j(P)={F_j(P)\over k-\chi_j}$$
In the standard Calogero-Moser
model, we have $l=1$, and $F_j=1$.

Let us summarize the results:

\begin{th}
 Let $\G$ be a curve that is defined by the equation of the form
(\ref{gammatrig}) or (\ref{gammarat}) and
$D=\gamma_1,\ldots,\gamma_{g+l-1}$ be a set of points in
general position. Then the formulas
\beq
{\rm Det ~} \Theta(x_i(t),t) =0
\eeq
\beq
a_{i, \alpha}(t)={1\over Q_i(t)}{\rm Det~} \Theta_{\alpha}^{(i)}, \quad
b_i^{\alpha}(t)={1\over Q_i^{+}(t)}{\rm Det~} \Theta^{+,\alpha (i)}
\eeq
where $Q_i(t)$ and $Q^+_i(t)$ are determined by the conditions
$f_{ii}=2$ and $\sum_\alpha b_i^\alpha =1$,
define the solutions of the system (\ref{4}, \ref{1.6}, \ref{1.7}).
Here $\Theta$ is an $N\times N$ matrix with elements given in
equations~(\ref{tetatrig}) in the trigonometric case and~(\ref{teta})
in the rational case. Moreover $\Theta_{\alpha}^{(i)}$ is obtained
from $\Theta$ by replacing its first column by $h_\alpha(T_j)$,
$j=1,\cdots,N$. Similarly for $\Theta^{+, \alpha ~(i)}$.
Any solution
of the system (\ref{1}) may be obtained from these solutions
taking into account the symmetries of the system.
\end{th}

\noindent {\Large \bf Appendix A}

\bigskip

\noindent The Weierstrass $\sigma$ function of periods
$ 2 \omega_1, 2 \omega_2 $ is the entire function defined by
\beq
  \sigma(z) = z \prod_{m,n\neq 0} \left ( 1 - \frac{z}{\omega_{mn}} \right )
  \exp \left [  \frac{z}{\omega_{mn}}
              + {1\over 2} \left ( \frac{z}{\omega_{mn}} \right ) ^2 \right ]
\eeq
with $ \omega_{mn} = 2 m \omega_1 + 2 n \omega_2 .$
The functions $\zeta$ and $\wp$ are
\beq
  \zeta(z) = \frac{\sigma'(z)}{\sigma(z)} \ , \ \ \ \ \ \wp(z) = -\zeta'(z),
\eeq
The $\wp$-function is doubly periodic, and the $\sigma$-function and
$\zeta$-functions transform
according to
\begin{eqnarray}
\zeta(z + \omega_l)= \zeta(z)+ \eta_l,\quad
\sigma(z+\omega_l)=-\sigma(z)e^{\eta_l (z+{\omega_l\over 2})}
\nonumber
\end{eqnarray}
where
\begin{eqnarray}
2(\eta_1 \omega_2 - \eta_2 \omega_1) = i\pi
\nonumber
\end{eqnarray}

These functions have the symmetries
\beq
  \sigma(-z) = - \sigma(z)\ , \ \ \ \ \ \zeta(-z) = - \zeta(z) \ , \ \ \ \ \
  \wp(-z) = \wp(z).
\eeq
Their behaviour at the neighbourhood of zero is
\beq
  \sigma(z) = z + O(z^5) \ , \ \ \ \ \  \zeta(z) = z^{-1} + O(z^3) \ ,
  \ \ \ \ \ \wp(z) = z^{-2} + O(z^2).
\eeq
Setting
\beq
  \Phi(x,z) =  \frac{\sigma(z-x)}{\sigma(x)\ \sigma(z)}e^{\zeta(z)x}
\eeq
it is easy to check that
\beq
  \label{a1}
  \Phi(-x,z) = - \Phi(x,-z) \ , \ \ \ \ \
 {d \over dx} \Phi(x,z) = \Phi(x,z) \ [\zeta(z+x) - \zeta(x)].
\eeq

The function $\Phi(x,z)$ is a double-periodic function of the variable $z$
\beq
\Phi(x,z+2\omega_l)=\Phi(x,z), \label{1.11a}
\eeq
and has the expansion of the form
\beq
\Phi(x,z)=(-z^{-1}+\zeta(x)+O(z))e^{\zeta(z)x} \label{1.11b}
\eeq
at the point $z=0$. As a function of $x$ it
has the following monodromy properties
\beq
\Phi(x+2\omega_l,z)=\Phi(x,z)\exp 2(\zeta(z)\omega_l-\eta_lz). \label{1.11d}
\eeq
and has a pole at the point $x=0$
\beq
\Phi(x,z)=x^{-1}+O(x), \label{1.11c}
\eeq

Choosing the periods $\omega_1=\infty$ and $\omega_2=i \frac{\pi}{2}$ we
obtain the hyperbolic case
\beq
  \label{tr1}
  \sigma(z)\to \sinh(z) \exp \left ( - \frac{z^2}{6} \right ) \ , \ \ \ \ \
  \zeta(z)\to \coth(z) - \frac{z}{3} \ , \ \ \ \ \
  \wp(z)\to \frac{1}{\sinh^2(z)} + \frac{1}{3}
\eeq
and
\beq
  \label{tr2}
  \Phi(x,z)\to \frac{\sinh(z-x)}{\sinh(z) \sinh(x)}
  e^{x\coth z}
\eeq
In the rational limit, we have
\begin{eqnarray}
\sigma(z) \to z, \quad \zeta(z) \to {1\over z}, \quad \wp(z)\to {1\over z^2},
\quad \Phi(x,z) \to \left( {1\over x} -{1\over z} \right)
e^{ {x\over z}  }
\nonumber
\end{eqnarray}

\noindent {\Large \bf Appendix B}

Let us recall briefly some facts we need about Riemann's
theta functions.

First there is an embedding of the Riemann
surface $\Gamma$ into its Jacobian $J(\Gamma)$ by the Abel map.

Let $a_i^0,b_i^0$ be a basis of cycles on $\Gamma$ with canonical matrix of
intersections $a_i^0\cdot a_j^0=b_i^0\cdot b_j^0=0, \ a_i^0\cdot
b_j^0=\delta_{ij}$. In a standard way it defines a basis of normalized
holomorphic differentials $\omega_j(P)$
\begin{equation}
\oint_{a_j^0} \omega_i=\delta_{ij}.
\end{equation}

The matrix of $b$-periods of these differentials
\begin{equation}
B_{ij}=\oint_{b_i^0}\omega_j
\end{equation}
defines the Riemann theta-function
\begin{equation}
 \theta(z_1,\ldots,z_g)=\sum_{m\in Z^g}e^{2\pi i(m,z)+\pi
i(Bm,m)}.
\end{equation}
on the torus  $J(\Gamma)$ which is called the Jacobian variety.
\begin{equation}
J(\Gamma)=C^g/{\cal B}
\end{equation}
The lattice ${\cal B}$ is generated
by the basic vectors $e_i\in C^g$ and by the vectors $B_j\in C^g$ with
coordinates $B_{ij}$.

The theta function has remarkable automorphy properties with respect
to this lattice: for any $l\in{\bf Z}^g$ and $z\in{\bf C}^g$
\begin{eqnarray}
\theta(z+l)&=&\theta(z)\nonumber\\
\theta(z+Bl) &=& \exp [-i\pi (Bl,l)-2i\pi (l,z)]\theta(z)\label{auto}
\end{eqnarray}

Let us choose a point $q_0\in \Gamma$. Then the vector $A(P)$ with coordinates
\begin{equation}
A_k(P)=\int_{q_0}^P \omega_k
\end{equation}
 defines the Abel map
\begin{equation}
A: \Gamma\longmapsto J(\Gamma)
\end{equation}
which is an embedding of $\Gamma$ into $J(\Gamma)$.

The fundamental theorem of Riemann expresses the intersection of the
image of this embedding with the zero set of the theta function.

\proclaim Theorem.
Let $Z=(Z_1,\cdots,Z_g)\in {\bf C}^g$ arbitrary. Either the function
$\theta(A(P)-Z)$ vanishes identically for $P\in\Gamma$ or it has
exactly $g$ zeroes $P_1,\cdots,P_g$ such that:
\begin{equation}
A(P_1)+\cdots+A(P_g)=Z-\cal K
\end{equation}
where $\cal K$ is the so--called vector of Riemann's constants,
depending on the curve $\Gamma$ and the point $q_0$ but independent of
$Z$.

 From this one can prove the Jacobi theorem, that is any point in the
Jacobian $J(\Gamma)$ is of the form $(A(P_1),\cdots,A(P_g)$ for some
divisor of degree $g$ on $\Gamma$.

This allows to find a formula for a function that has
$g$ poles at points $\delta_1,\ldots, \delta_g$ and
an additional pole at point $Q^+$. The dimension of the space of
such  functions is two. Of course it contains the constant. We
choose the second basic function by the condition that it vanishes
at some fixed point $Q^-$.

Let $Z,Z^+,Z^-,Z^0$ be vectors that are defined by formulae:
\begin{eqnarray*}
Z&=&\sum_{s=1}^g A(\delta_s)+{\cal K}\\
Z^+&=&Z-A(\delta_1)+A(Q^+)=A(Q^+)+\sum_{s=2}^g A(\delta_s)+{\cal K}\\
Z^-&=&Z-A(\delta_1)+A(Q^-)\\
Z^-+Z^0&=&Z+Z^+
\end{eqnarray*}
Let us define the function
\begin{equation}
f(P)={\theta(A(P)-Z^-)\theta(A(P)-Z^0)\over
\theta(A(P)-Z) \theta (A(P)-Z^+)} \label{laformule}
\end{equation}

 From the Jacobi theorem it follows that two factors in the denominator
vanish at the points $\delta_1,\ldots,\delta_g$ and $Q^+,\delta_2,\ldots,
\delta_g$ respectively. Similarly the two factors in the numerator
vanish at $Q^+,\delta_2,\ldots,\delta_g$ and $g$ other points.

The zeroes at $\delta_2,\ldots,\delta_g$
cancel between the numerator and the denominator, thereby leaving us
with the correct divisor of zeroes and poles. It remains to show that
the function $f$ is well--defined on $\Gamma$. This is because, due to
the definition of $Z^0$ the automorphy factors of the theta functions
in equation~(\ref{auto})
cancel between the numerator and the denominator when $P$ describes
$b$--cycles on $\Gamma$.

\section{Acknowledgements}
One of the authors I.K. would like to thank
the Institut Henri Poincar\'e for hospitality during the time
when this work had been done. This work was supported by RFFI grant
93-011-16087 and ISF grant MD800.

\end{document}